\documentclass[pra,floatfix,footinbib,superscriptaddress, twocolumn, amsmath,amssymb,amsfonts]{revtex4-1} 

\usepackage[T1]{fontenc}
\usepackage[latin9]{inputenc}
\usepackage{lmodern} 
\usepackage{color}
\usepackage{bbold}
\usepackage{dsfont}
\usepackage{graphicx}
\usepackage[caption=false]{subfig}
\usepackage[colorlinks]{hyperref}
\usepackage[bottom]{footmisc}
\usepackage{enumerate}
\usepackage{float}
\usepackage{mathtools}
\usepackage{leftindex}

\usepackage{empheq}
\usepackage{tikz}
\usepackage[normalem]{ulem}
\DeclareMathAlphabet\mbc{OMS}{cmsy}{b}{n}
\usepackage{balance}

\begin{document}

\global\long\def\eqn#1{\begin{align}#1\end{align}}
\global\long\def\vec#1{\overrightarrow{#1}}
\global\long\def\ket#1{\left|#1\right\rangle }
\global\long\def\bra#1{\left\langle #1\right|}
\global\long\def\bkt#1{\left(#1\right)}
\global\long\def\sbkt#1{\left[#1\right]}
\global\long\def\cbkt#1{\left\{#1\right\}}
\global\long\def\abs#1{\left\vert#1\right\vert}
\global\long\def\cev#1{\overleftarrow{#1}}
\global\long\def\der#1#2{\frac{{d}#1}{{d}#2}}
\global\long\def\pard#1#2{\frac{{\partial}#1}{{\partial}#2}}
\global\long\def\re{\mathrm{Re}}
\global\long\def\im{\mathrm{Im}}
\global\long\def\dd{\mathrm{d}}
\global\long\def\ddd{\mathcal{D}}
\global\long\def\hmb#1{\hat{\mathbf #1}}
\global\long\def\avg#1{\left\langle #1 \right\rangle}
\global\long\def\mr#1{\mathrm{#1}}
\global\long\def\mb#1{{\mathbf #1}}
\global\long\def\mc#1{\mathcal{#1}}
\global\long\def\tr{\mathrm{Tr}}
\global\long\def\dbar#1{\Bar{\Bar{#1}}}

\global\long\def\nth{$n^{\mathrm{th}}$\,}
\global\long\def\mth{$m^{\mathrm{th}}$\,}
\global\long\def\non{\nonumber}

\newcommand{\orange}[1]{{\color{orange} {#1}}}
\newcommand{\cyan}[1]{{\color{cyan} {#1}}}
\newcommand{\blue}[1]{{\color{blue} {#1}}}
\newcommand{\yellow}[1]{{\color{yellow} {#1}}}
\newcommand{\green}[1]{{\color{green} {#1}}}
\newcommand{\red}[1]{{\color{red} {#1}}}
\newcommand{\ks}[1]{{\color{orange}[KS: {#1}]}}
\newcommand{\ad}[1]{{\color{red}[AD: {#1}]}}
\global\long\def\todo#1{\orange{{$\bigstar$ \cyan{\bf\sc #1}}$\bigstar$} }

\title{Non-Markovian spontaneous emission in a tunable cavity formed by atomic mirrors}

\author{Annyun Das}
\email{annyun@arizona.edu}
\affiliation{Wyant College of Optical Sciences and Department of Physics, University of Arizona, Tucson, AZ 85721}
\author{Pablo Solano}
\email{psolano@udec.cl}
\affiliation{Departamento de F\'{i}sica, Facultad de Ciencias F\'{i}sicas y Matem\'{a}ticas, Universidad de Concepci\'{o}n, Concepci\'{o}n,
Chile}
\author{Kanu Sinha}
\email{kanu@arizona.edu}
\affiliation{Wyant College of Optical Sciences and Department of Physics, University of Arizona, Tucson, AZ 85721}

\begin{abstract}
We analyze the non-Markovian spontaneous emission dynamics of a two-level test atom placed in a cavity formed by two atomic arrays in a waveguide quantum electrodynamics (QED) setup. We demonstrate a crossover from single-mode to multimode strong coupling cavity QED as the cavity length $ \sim d$ becomes comparable to the  coherence length associated with collective spontaneous emission $\sim v/(N\gamma)$. The resulting non-Markovian dynamics of the test atom and the emergent spectral density of the  field are analyzed as a function of various tunable atomic array parameters: number of atoms, length of the atomic cavity, and resonance frequency of the atoms forming the atomic mirrors. 
Our results show limitations to  cooperatively enhanced light-matter coupling  in  the presence of time-delayed feedback.  We further illustrate that the non-Markovian system dynamics can be efficiently approximated in terms of a few modes of the emergent spectral density of the field. 


 \end{abstract}

\maketitle

\section{Introduction}

As a collection of $N$ atoms interacts coherently with the electromagnetic (EM) field, the interference between scattered fields from individual atoms can lead to a cooperatively enhanced light-matter coupling $ \sim \sqrt{N} g$ (where $g$ is the single atom-photon coupling strength)~\cite{Dicke1954, Gross1982,Guerin2017,Hammerer2010}. Such enhancement is central to a range of phenomena  from Dicke superradiance~\cite{Wang2007,Goban2015}, to realization of strong light-matter interaction~\cite{Chen2010,Chang2012,Ruddell2017,Li2023}, efficient transfer of quantum information~\cite{Duan2001,Sangouard2011},  and superabsorption of light in molecular complexes~\cite{Higgins2014,Yang2021}. As a remarkable feature, a spatially ordered collection of atoms interacting cooperatively with the EM field mode can  alter its  mode structure: Atomic arrays behave as mirrors for near-resonant field modes~\cite{Shen05, Chang2012, Weidemuller1995, Birkl1995, Corzo2016,Sorensen2016}. 

Notably, such atomic mirrors not only exhibit a cooperatively enhanced light-matter coupling themselves but can also lead to strong coupling dynamics  for a nearby `test' atom~\cite{Chang2012,Guimond2016}. Such amplified secondary cooperative coupling can be attributed to the modification of the effective spectral density of the EM field by the atomic mirror. For example, in the case of a test atom placed in a `cavity' region formed by two atomic mirrors with $N$ atoms each, the emergent coupling between the test atom and its environment goes as $ \sim \sqrt{N} \gamma$~\cite{Chang2012} ($\gamma$ being the emission rate of the test atom). This leads to the question: how is this secondary cooperative coupling and the dynamics of the test atom impacted by an emergent multimode spectral density of the field?

\begin{figure}[b]
     \centering
     \includegraphics[width = 0.45\textwidth]{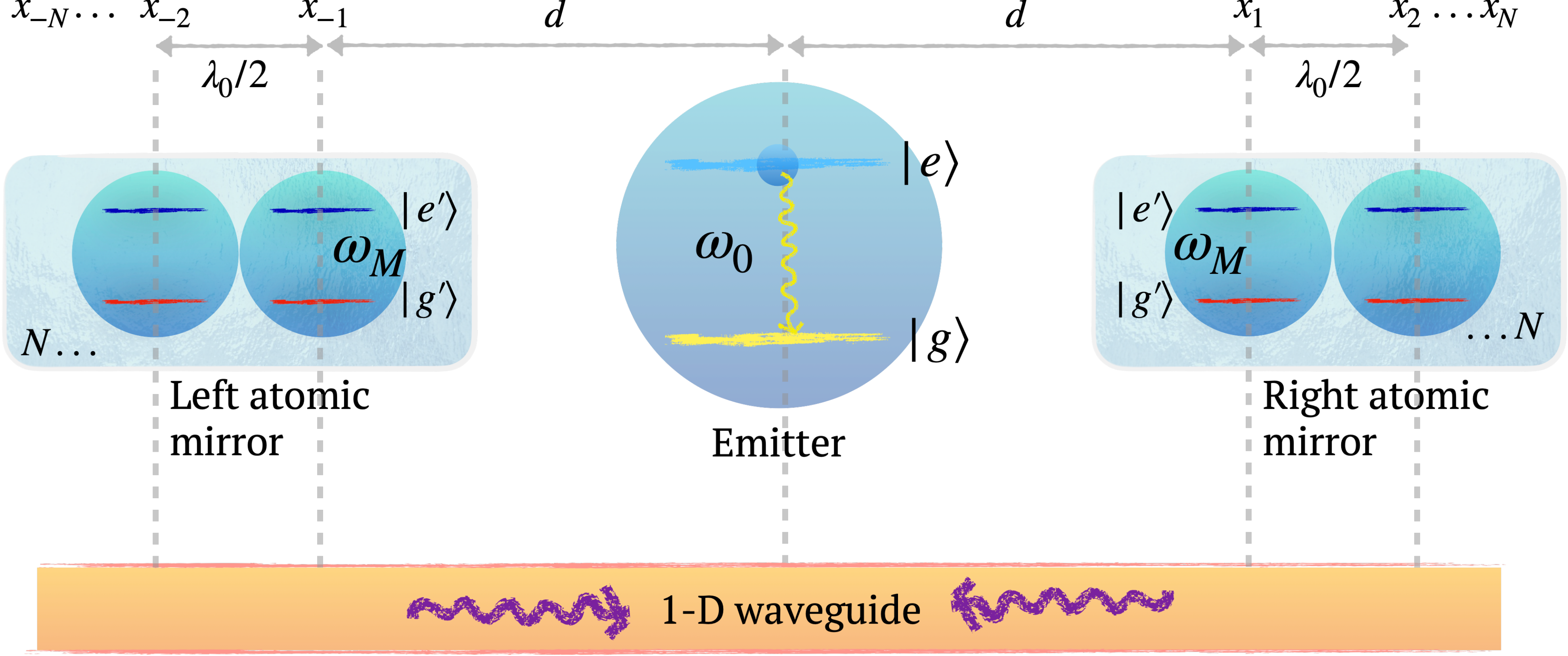}
     \caption{ Two arrays of $N$ two-level atoms of resonant frequency $\omega_M$ between $\ket {g'}$ and $\ket {e'}$ and nearest neighbor separation $\lambda_0/2=\pi v/\omega_0$ coupled to a one-dimensional waveguide. An excited  two-level emitter with resonant frequency $\omega_0$ is placed in between the two arrays at a distance $d$ from each array.  }
     \label{Fig:CavitySchematic}
 \end{figure}
We investigate this question in a waveguide quantum electrodynamics (QED) setup with a test atom  in a `cavity' formed by two one-dimensional atomic Bragg mirrors (see Fig.\,\ref{Fig:CavitySchematic}). As waveguides enable direct coupling of emitters separated by large distances~\cite{Sheremet23}, such a setup offers a promising path to combine strong-coupling QED with long-ranged interactions~\cite{Guimond2016}. For sufficiently large separations between the atomic mirrors,  as the cavity free spectral range ($\sim v/d$) becomes comparable to the effective linewidth of the test atom ( $\sim N\gamma$), the  system exhibits rich multimode dynamics \cite{Giessen1996, Krimer2014} ($v$ being the speed of light in the waveguide). Concomitantly, as the cavity-round trip time ($\sim d/v$) becomes comparable to the atomic relaxation rate ($\sim 1/(N\gamma)$), the non-Markovian time-delayed feedback from the field on the test atom becomes pertinent~\cite{Giessen1996,Sinha20a, Sinha20b}.  We analyze the interplay between such multimode dynamics (varying the cavity length $d$), and the enhanced secondary cooperative coupling due to the atomic mirrors (varying the number of atoms in each atomic mirror $N$).  We demonstrate a crossover from single-mode to  multimode strong coupling regimes in terms of the resulting non-Markovian spontaneous emission  dynamics of the test atom and  the effective spectral density of the field.  Such a spectral analysis of the system dynamics is critical for keeping track of the energy and information distributed across multiple field modes, and allows for efficient computation of non-Markovian system dynamics pertinent to various long-distance waveguide QED protocols~\cite{Crowder2024, Zheng, Giron}. Finally, we examine the  dynamics and spectral response of the system in the presence of detuning between the test atom and the nearest cavity mode, as well as the resonance frequency of the individual atoms in the atomic mirrors.


\section{Model}
Our setup consists of a two-level `test' atom, which we will refer to as the `emitter',  placed in between two arrays of two-level atoms coupled via a one-dimensional waveguide as shown in Fig.\,\ref{Fig:CavitySchematic}. The emitter, with ground and excited levels denoted by $ \ket{g}$ and $\ket {e}$, has a resonance frequency $\omega_0$. The array atoms, with $\ket {g^\prime}$ ($\ket {e^\prime}$) as the ground  (excited) levels,  are arranged at distances of $\lambda_0/2$ from each other, where $\lambda_0=2\pi v/\omega_0$. In this configuration, the arrays behave as Bragg mirrors for an incident field of wavelength $ \lambda_0$, and the two arrays constitute a cavity~\cite{Shen05, Chang2012, Guimond2016}. The array atoms have a resonant frequency $\omega_M = \omega_0 - \delta$, which is detuned by $\delta$ relative to the emitter's resonance frequency. 
The right-most and left-most atoms of the left and right atomic-mirrors are placed at $ x_{-1} = -d$ and $x_1 = d$, respectively.

The total Hamiltonian for the system is $\hat H = \hat{H}_0+\hat{H}_\mr{int}$, with $ H_0$ as the free Hamiltonian:
\eqn{\hat H_0  = {\sum_{n=-N}^N\hbar\omega_n\hat{\sigma}_n^+\hat{\sigma}_n^-+\int _{-\infty}^\infty dk\hbar\omega \hat{a}_k^\dagger\hat{a}_k}.} Here  $\hat \sigma_n^\pm$ are the  ladder  operators  corresponding to the $n^\mr{th}$ atom: $n=0$ being the emitter, indices $n\in[-N,-1]$ refer to the atoms in the left array, and $n\in[1,N]$ to the atoms in the right array. The array atoms have a resonance frequency $\omega_n=\omega_M $ ($\forall n\neq 0$).  $\cbkt{\hat{a}^\dagger_k, \hat{a}_k}$  represent the creation and annihilation operators for the guided modes with wavevector $k = \omega/v$, assuming linear dispersion. $\hat{H}_\mr{int}$ is the electric-dipole interaction Hamiltonian between the atoms and the waveguide field modes, under the rotating-wave approximation (RWA), which can be expressed in the interaction picture as $\tilde H_\mr{int} \equiv e^{i  \hat{H}_0t}\hat{H}_\mr{int}e^{-i  \hat{H}_0t}$: 
\eqn{\label{CavityHamiltonian}
\tilde{H}_{int} = 
{\sum_{n=  -N}^N \int_{-\infty}^\infty dk \hbar g_k \hat a_k \hat \sigma_n ^+   e ^{i  k x_n }  e^{-i  \bkt{\omega - \omega _n }t }+ \mr{H. c.}}
}

The initial state of the total atom+field system is assumed to be $\ket {\Psi(0)}= \ket e \otimes\ket{G_L'G_R'}\otimes\ket {\cbkt{0}}$ where $\ket {G_L^\prime}\equiv\ket {g_{-1}^\prime}\otimes\ket {g_{-2}^\prime}...\otimes\ket {g_{-N}^\prime}$, $\ket {G_R^\prime}\equiv\ket {g_{1}^\prime}\otimes\ket {g_{2}^\prime}...\otimes\ket {g_{N}^\prime}$  and $\ket {\cbkt{0}}$ is the vacuum state of the EM field. Considering that the Hamiltonian preserves the total number of excitations for the atom+field system,
the state at a general time $t$ is given by:
\eqn{
\ket {\Psi(t)} =& \sbkt{c_0(t)\hat\sigma_0^++\sum_{n=-1}^{-N}c_{e_n,G'}(t)\hat\sigma_n^++\sum_{n=1}^{N}c_{G',e_n}(t)\hat\sigma_n^+\right.\non\\
&\left.+\int_{-\infty}^\infty dk c_{k}(t)\hat a_k^\dagger}\ket{G_L'G_R'}\ket g \ket{\cbkt 0}.\label{CavityAnsatz}
}    
 $c_0(t)$ is the excitation  amplitude of the emitter and $c_{e_n,G'}(t)$ ($c_{G',e_n}(t)$) is the amplitude for the $n^\mr{th}$ atom in the left (right) array. $c_{k}(t)$ is the excitation probability amplitude of field mode with wavevector \textit{k}.

\subsection{System dynamics}
   We solve the Schr\"odinger equation in the interaction picture and by tracing over the field modes, we obtain the following coupled delay differential equations (DDEs) for the excitation amplitudes of the emitter and atomic mirrors (see Appendix~\ref{AppendixA} for details):


\begin{widetext}
\begin{subequations}
    \begin{align}
        \frac{dc_0(t)}{dt}= &-\frac{\gamma}{2} \sbkt{c_0(t) +\sqrt{N}c_{LM}\bkt{t-\frac{d}{v}}e^{i\phi_M}e^{i\delta t}\Theta\bkt{t-
        \frac{d}{v}}+\sqrt{N}c_{RM}\bkt{t-\frac{d}{v}}e^{i\phi_M}e^{i\delta t}\Theta\bkt{t-\frac{d}{v}}}\label{CavityDDEa} \\
        \frac{dc_{LM}(t)}{dt}= &-\frac{\gamma}{2}\sbkt{\sqrt{N}c_0\bkt{t-\frac{d}{v}}e^{-i\delta t}e^{i\phi_0}\Theta\bkt{t-\frac{d}{v}}+Nc_{LM}(t)+Nc_{RM}\bkt{t-\frac{2d}{v}}e^{2i\phi_M}\Theta\bkt{t-\frac{2d}{v}}}\label{CavityDDEb} \\
        \frac{dc_{RM}(t)}{dt}= &-\frac{\gamma}{2}\sbkt{\sqrt{N}c_0\bkt{t-\frac{d}{v}}e^{-i\delta t}e^{i\phi_0}\Theta\bkt{t-\frac{d}{v}}+Nc_{RM}(t)+Nc_{LM}\bkt{t-\frac{2d}{v}}e^{2i\phi_M}\Theta\bkt{t-\frac{2d}{v}}}\label{CavityDDEc}
    \end{align}
\end{subequations}
\end{widetext}
The $\lambda_0/2$ lattice spacing in each atomic mirror allows us to simplify the coupled atomic dynamics by defining collective excitation amplitudes $c_{LM}(t)= \sum_{j=-1}^{-N} c_{e_j,G^\prime}(t)e^{(j+1)i  \pi}/\sqrt N$ for the left array and $c_{RM}(t)= \sum_{j=1}^{N} c_{G^\prime,e_j}(t)e^{(j-1)i  \pi}/\sqrt N$ for the right array which couple to the guided EM field with an enhanced coupling of $\sqrt{N}g_0$. $\gamma\equiv2\pi|g_0|^2$ is the rate of spontaneous emission of an individual atom into the 1-D waveguide, assuming a flat spectral density, such that $g_k\approx g_0$ in the spectral region around the atomic resonance. We identify the propagation phases $\phi_{0(M)}=\omega_{0(M)}d/v$ for fields emitted from the emitter (atomic mirrors). We assume a perfect coupling between the atoms and the waveguide.

The dynamics of the emitter is affected by that of both mirrors with a delay of \textit{d/v} (Eq.\,(\ref{CavityDDEa})) and vice versa, and the arrays influence each other's evolution with a delay of 2\textit{d/v} (Eqs. \eqref{CavityDDEb}, \eqref{CavityDDEc}). In the regime where the dimensionless distance $\eta\equiv\gamma d/v$ is significant $ (\eta\sim 1)$, the relaxation timescale for the emitter $(\sim1/\gamma)$  becomes comparable to the time the field takes to mediate information between the emitter and the arrays $(\sim d/v)$, bringing non-Markovian time-delayed feedback into consideration ~\cite{DornerZoller, Sinha20a, Sinha20b, Dinc2019}. The emitter exhibits an enhanced coupling to its environment ($ \sim \sqrt{N} \gamma$), corresponding to the collective excitations in the atomic arrays~\cite{Chang2012}. Additionally, the detuning between the emitter and the mirror introduces a dynamic phase $e^{\pm i \delta t}$ in the emitter and atomic-mirrors' evolutions, resulting in beats. The symmetry between Eqs. \eqref{CavityDDEb} and \eqref{CavityDDEc} suggests that the left and right arrays evolve in an identical fashion with time for the given initial state.

The field intensity is obtained as $I = {\epsilon_0v}\bra{\Psi(t)}\hat{E}^\dagger(x,t)\hat{E}(x,t)\ket{\Psi(t)}$.  The electric field  operator is given by $\hat{E}\bkt{x,t}= $ $\int_{-\infty}^\infty dk {E}_k \hat{a}_ke^{i  k x}e^{- i \omega t}$, with $E_k=i\bkt{\frac{\hbar\omega }{\pi\epsilon_0A}}^{1/2}$ (\textit{A} being the cross-section of the waveguide) \cite{Blow} which yields  (see Appendix~\ref{AppendixIntensity} for details)
\begin{widetext}
\eqn{\label{Eq:FieldIntensity}
\frac{I(x,t)}{I_0}=&\left|c_0\bkt{t-\frac{x}{v}}e^{-i\omega_0(t-x/v)}\cbkt{\Theta\bkt{t-\frac{x}{v}}-\Theta\bkt{-\frac{x}{v}}}+c_0\bkt{t+\frac{x}{v}}e^{-i\omega_0(t+x/v)}\cbkt{\Theta\bkt{t+\frac{x}{v}}-\Theta\bkt{\frac{x}{v}}}\right.\non\\
&\left.+\sqrt N\left[c_{LM}\bkt{t-\frac{x+d}{v}}e^{-i\omega_M(t-(x+d)/v)}\cbkt{\Theta\bkt{t-\frac{x+d}{v}}-\Theta\bkt{-\frac{x+d}{v}}}\right.\right.\non\\
&\left.\left.+c_{LM}\bkt{t+\frac{x+d}{v}}e^{-i\omega_M(t+(x+d)/v)}\cbkt{\Theta\bkt{t+\frac{x+d}{v}}-\Theta\bkt{\frac{x+d}{v}}}\right]\right.\non\\
&\left.+\sqrt N \left[c_{RM}\bkt{t-\frac{x-d}{v}}e^{-i\omega_M(t-(x-d)/v)}\cbkt{\Theta\bkt{t-\frac{x-d}{v}}-\Theta\bkt{-\frac{x-d}{v}}}\right.\right.\non\\
&\left.\left. +c_{RM}\bkt{t+\frac{x-d}{v}}e^{-i\omega_M(t+(x-d)/v)}\cbkt{\Theta\bkt{t+\frac{x-d}{v}}-\Theta\bkt{\frac{x-d}{v}}}\right]\right|^2,
}
\end{widetext}
as the intensity, where $I_0 = \frac{\epsilon_0 \gamma \abs{E_{k_0}}^2}{2\pi}$ . The first two terms correspond to the right- and left-going field modes respectively, emanating from the emitter, placed at $ x_0 = 0$. Similarly, the third and fourth (fifth and sixth) terms  represent the right- and left-going modes from the left (right) atomic-mirror at $x_{-1} = -d $ ($x_1=d$).
\subsection{Frequency response}

The frequency response
 of the emitter can be analyzed by decoupling the DDEs Eq.\,\eqref{CavityDDEa}-\eqref{CavityDDEc} in the frequency domain to yield  (see Appendix~\ref{Appendix:ResponseFunctions} for details):
\eqn{\label{Eq:FormalFourierInverseEmitter}
&c_0(t) = \int_{-\infty}^{\infty}d\omega F_0(\omega)e^{-i \omega t}
}
with
  
\eqn{\label{EmitterResponseFunction}
        &F_0(\omega)\equiv \frac{1}{2\pi i}\frac{1}{D_0 \bkt{\omega}} }
        as the Fourier transform of the emitter's excitation amplitude, and its denominator: 
\eqn{\label{Eq:DenominatorNoTaylor}
&D_0\bkt{\omega}\equiv\non\\
&\frac{\gamma}{2}- i  \omega-\frac{N\gamma^2e^{2 i  \phi_0}}{e^{2 i \phi_0}N\gamma +e^{-2 i \omega d/v}\bkt{N\gamma-2 i \omega-2i\delta}}.
}

\begin{figure*}[ht]
   \includegraphics[width = 1\linewidth]{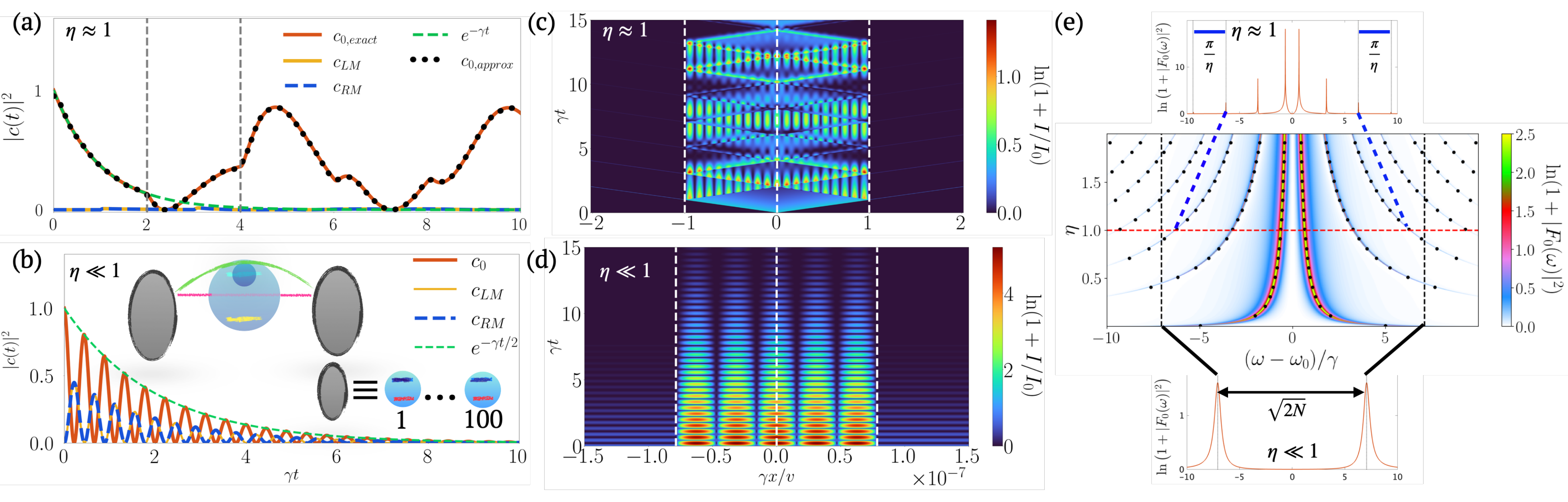}
    \caption{  Atomic and field dynamics, and the spectral density  for $\phi_0$ a half-integer multiple of $\pi$ (emitter at cavity antinode). We assume $\delta=0$ and $N=100$ atoms in each array.(a) Non-Markovian spontaneous emission dynamics for $\eta \approx1$. The $c_{0,exact}$, $c_{0,approx}$, $c_{LM}$, $c_{RM}$ curves refer to the excitation probabilities for the emitter plotted using Euler's method, by using Eq.\,\eqref{Eq:ApproximateEmitterDynamicsNM}, and the excitation probabilities of the left and right atomic mirrors respectively. (b) Damped Rabi oscillations   in the single mode regime $(\eta\ll1)$. Schematic shows that the emitter is placed at an antinode between the two atomic mirrors composed of $N=100$ atoms.  (c) The field intensity as a function of position and time for $\eta\approx1$. The dashed vertical lines represent the positions of the emitter  $(x=0)$ and the atomic arrays (at $-\eta$ and $\eta$). (d) The field intensity for $\eta\ll1$. (e) Semi-logarithmic plot of the spectral density $\abs{F_0\bkt{\omega}}^2$ as a function of $\omega$ and $\eta\in \cbkt{0,2}$. The black dots represent the characteristic frequencies $ \omega_p$ obtained via  Eq.\,\eqref{Eq:DerivativeDenominator}. Inset below corresponding to $\eta\ll1$ shows the frequency peaks with a vacuum Rabi splitting in the single-mode cavity regime. Inset above corresponding to the red horizontal line at $ \eta \approx 1$ shows the multiple frequency peaks observed in the multimode regime.  }
    \label{Fig:AntiNodeNoDetuningCavity}
\end{figure*}

To determine the characteristic frequencies present in the dynamics, and the corresponding peaks in the spectral density, we minimize  the absolute value of the $ D_0 \bkt{\omega}$ (see Appendix~\ref{Appendix:ResponseFunctions} for details):
\eqn{\label{Eq:DerivativeDenominator}
\left.\der{}{\omega} \abs{D_0 \bkt{\omega}}\right\vert_{\omega_p}=0.} 
This yields the characteristic frequencies $ {\omega_p} $ present in the emitter dynamics, centered around $ \omega_0$, as we will illustrate in the following section. 

In the limit of $ N\eta\ll1$, when $\delta=0$, the above condition yields the characteristic frequencies:
\eqn{\label{Eq:omegap}\omega_p\approx&\pm\frac{\sqrt{-1+4N}\gamma}{2\sqrt 2}; \quad\text{for } \phi_0=\pi/2\\
\omega_p = & ~0; \quad \text{for } \phi_0 = \pi.
}
\begin{figure}[b]
   \includegraphics[width = 1\linewidth]{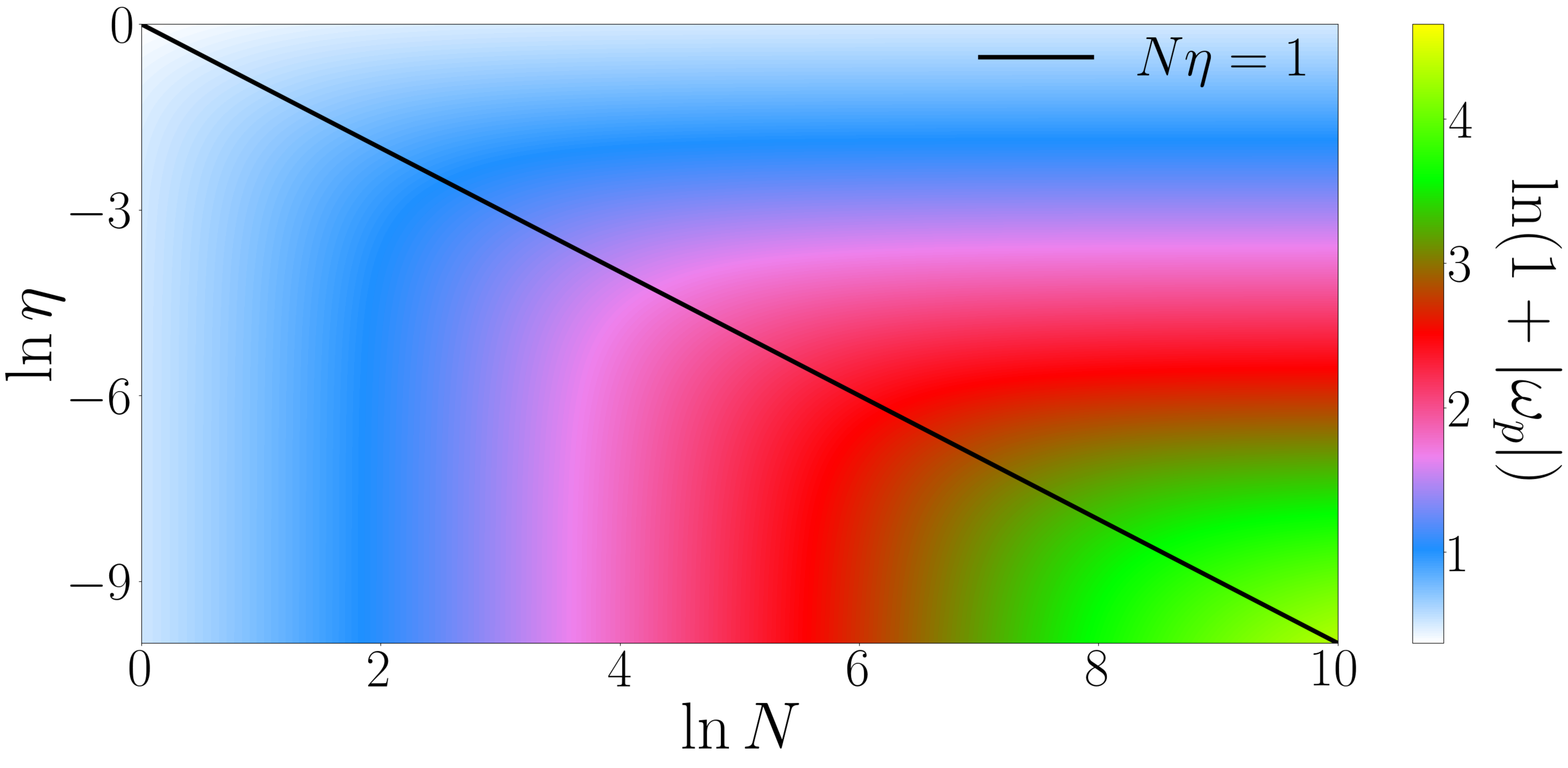}
    \caption{ Absolute value of the first characteristic frequencies $\omega_p$ to the left and right of $\omega=0$ for a range of N and $\eta$ for $\delta=0$ when the emitter is placed at an antinode.}
    \label{Fig:FrequencyResponseEtaRange}
\end{figure}
 As $ N\eta \sim 1$, the cavity formed by the two atomic arrays has resonant frequencies separated by the  free spectral range (FSR) $\Delta \omega_c/\gamma \equiv \pi/2\eta$,  with an effective mode structure established between the two atomic mirrors such that the emitter can be positioned at a node or an antinode of the cavity by adjusting the separation $d$. Placing the emitter in the middle allows it to act as a membrane that splits the cavity into two half-cavities with FSR of $\pi/\eta$ each.  As we will see below the spontaneous emission spectrum of the emitter can be understood in terms of the resonances of the effective  cavity and the half-cavities.  As the cavity length becomes long such that the effective FSR is comparable to the emitter's linewidth, multiple modes of the cavity become pertinent, bringing non-Markovian time-delayed feedback into consideration~\cite{Giessen1996}.  In the following section we will further demonstrate that the position of the emitter strongly influences the emergent spectral density of the field modes.

 The emitter's dynamics can be well-approximated via the sum of the residues  by using Cauchy's Residue theorem in Eq.\,\eqref{Eq:FormalFourierInverseEmitter} for the most pertinent characteristic frequencies in the emitter's dynamics. The complex poles $\tilde\omega_p$ of Eq.\,\eqref{EmitterResponseFunction} found by setting $D_0(\omega)=0$ closely approximate $\omega_p$ (Eq.\,\eqref{Eq:DerivativeDenominator}).  
 This allows us to write an approximate solution to the emitter's dynamics as 
 \eqn{\label{Eq:ApproximateEmitterDynamicsNM}
 c_0(t)~\approx ~\lim_{\omega\to\tilde\omega_p}\sum_{p = -P }^P\frac{\omega-\tilde \omega_p}{D_0(\omega)},
 }
 where  we consider $P$ poles around $ \omega_0 $.
\begin{figure*}[t]
   \includegraphics[width = 1\linewidth]{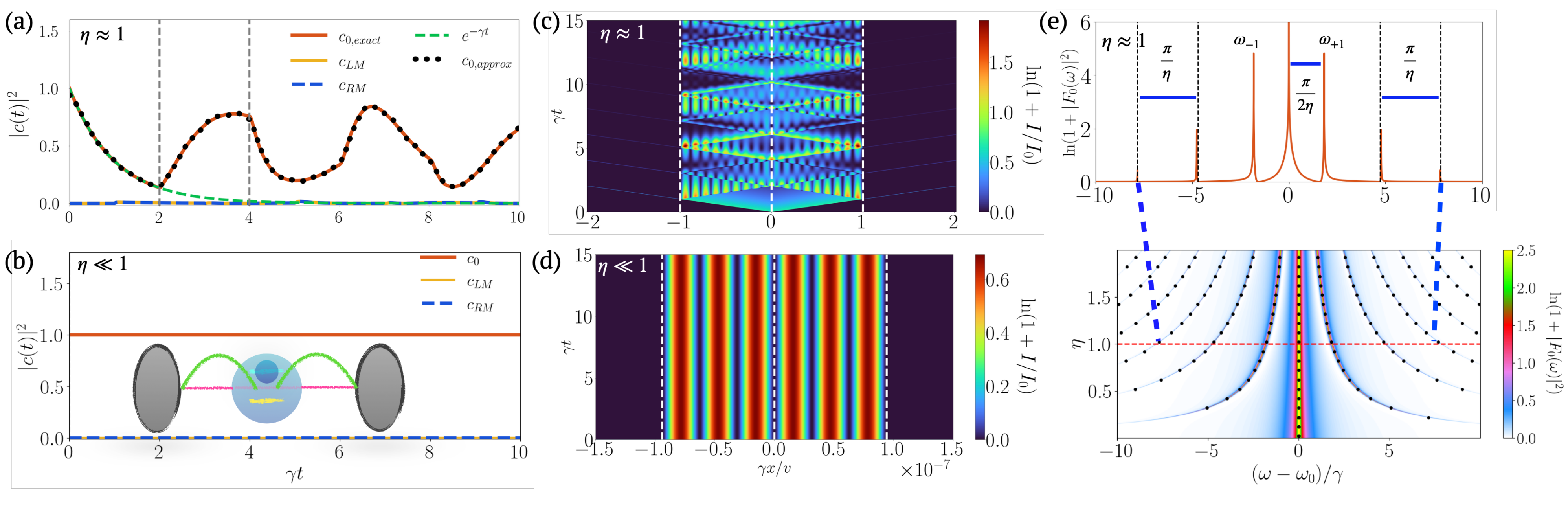}
    \caption{   Atomic and field dynamics, and the spectral density  for $\phi_0$ an integer multiple of $\pi$ (emitter at a node). We assume $\delta=0$ and $N=100$ atoms in each array. (a) Non-Markovian spontaneous emission dynamics for $\eta \approx 1$. The $c_{0,exact}$, $c_{0,approx}$, $c_{LM}$, $c_{RM}$ curves refer to the excitation probabilities for the emitter plotted using Euler's method, by using  Eq.\,\eqref{Eq:ApproximateEmitterDynamicsNM}, and the excitation probabilities of the left and right arrays respectively.  (b) Atomic dynamics  for the emitter and atomic arrays for $\eta\ll1$.  The schematic shows the emitter placed at a node between the two arrays. (c) The field intensity as a function of position and time in the multimode regime $( \eta \approx 1)$. (d) The field intensity in the single-mode cavity regime. (e) Semi-logarithmic plot of the spectral density $\abs{F_0\bkt{\omega}}^2$ as a function of $\omega$ and $\eta\in \cbkt{0,2}$. The black dots represent the characteristic frequencies $ \omega_p$ obtained via  Eq.\eqref{Eq:DerivativeDenominator}. The inset shows the frequency spectrum for $\eta\approx1$.}
    \label{Fig:NodeNoDetuningCavity}
\end{figure*}
\section{System dynamics and frequency response: Single-mode to multimode crossover}
We now consider the system dynamics going from a short single-mode cavity $\bkt{N\eta\ll1}$ to long  multimode cavity $\bkt{N\eta\gtrsim1}$ regime,   with an initially excited emitter positioned at either the node ($\phi_0=n\pi$) or antinode ($\phi_0=(2n+1)\pi/2$) of the cavity formed by the atomic mirrors. We assume here that the detuning between the emitter and the atoms forming the mirrors is $ \delta = 0$. 

\subsection{Emitter at cavity antinode}     Fig.\,\ref{Fig:AntiNodeNoDetuningCavity}\,(b) shows that when $\eta\ll1 $ and $ N\eta\ll1$, the emitter,  initially prepared in the excited state and placed at an antinode of the cavity, undergoes damped Rabi oscillations as is well-established in cavity QED \cite{MeystreBook2007,Chang2012,Guimond2016}. The atomic mirrors also undergo damped Rabi oscillations in this regime~\cite{Sinha2024}. We observe such Rabi oscillations in the emitted intensity  in Fig.\,\ref{Fig:AntiNodeNoDetuningCavity}(d)  as the single excitation is exchanged between the atoms and the field modes  until the excitation is lost to the field modes outside the cavity.

For $\eta\approx1$ $(N\eta\gg1)$, the emitter exhibits complex multi-mode dynamics (Fig.\,\ref{Fig:AntiNodeNoDetuningCavity}(a)), with a non-Markovian time-delayed feedback from the field at intervals of $\Delta t=2d/v$ (vertical dashed-gray lines). At $t={2d}/{v}$, the emitter's spontaneous emission dynamics deviates from exponential decay $e^{-\gamma t}$ (green-dashed curve).  From the field intensity plot Fig.\,\ref{Fig:AntiNodeNoDetuningCavity}(c), we note that the light emitted at $t=0$ gets reflected from the atomic mirrors at $t=d/v$ and an interference pattern emerges in between the two atomic mirrors. 
The single excitation is exchanged between the emitter, the arrays, and the cavity field modes without significant loss to the outside field modes.
In contrast, in the  single-mode cavity regime ($ \eta\ll 1$), the interference pattern is established with no time delay and the single excitation is  lost to outside field modes exponentially as $\sim \text{exp}\bkt{-\gamma t/2}$ (Fig.\,\ref{Fig:AntiNodeNoDetuningCavity}(d))\cite{Sinha2024}.

Fig.\,\ref{Fig:AntiNodeNoDetuningCavity}(e) shows the  spectral density $ \abs{F_0 \bkt{\omega}}^2$ (Eq.\eqref{EmitterResponseFunction}). Considering a specific value of $ \eta$, the intersection of horizontal line $ \eta=\eta_0$ with the spectral density peaks informs us which frequencies contribute to the emitter's dynamics. The $c_{0,approx}$ curve in Fig.\,\ref{Fig:AntiNodeNoDetuningCavity}(a) shows the emitter's dynamics approximated using Cauchy's residue theorem for the first three frequencies from each side of $\omega=0$ in Eq.\,\eqref{Eq:ApproximateEmitterDynamicsNM}. For ${\eta\lesssim10^{-4}}$ $(N\eta\ll1)$, the zeroth order term in $\eta$ in Eq.\,\eqref{Eq:DenominatorNoTaylor} gives us two poles separated by the Rabi splitting of $\approx\sqrt{2N}\gamma$ for $N\gg1$ (Fig.\,\ref{Fig:AntiNodeNoDetuningCavity}(e) bottom inset). This is a consequence of the emitter's strong coupling with the effective cavity modes ~\cite{Agarwal1984}.

For larger $ \eta$ values, multiple characteristic frequencies appear in the dynamics, which can be attributed to the resonances of the effective cavity formed by the two atomic mirrors. As seen in Fig.\,\ref{Fig:AntiNodeNoDetuningCavity}(e) for  $\eta\approx 1$ multiple peaks appear, separated by the half-cavity FSR of $\pi/\eta$. This corresponds to the frequencies that exhibit antinodes at the position of the emitter. With large values of $N$, the collective coupling strength $N\gamma$ can exceed the FSR, facilitating the `superstrong' coupling regime of cavity QED~\cite{Meiser2006, Krimer2014, Johnson2019, Blaha2022}. The intersection of the horizontal dashed-red line with the peaks of $ \abs{F\bkt{\omega_0}}^2$ shows the contributing frequencies at $\eta\approx1$. 
By Taylor expanding the solutions to Eq.\,\eqref{Eq:DerivativeDenominator}, we obtain $\omega_p$ to $\mathcal O[\eta]^3$ as:
\eqn{\label{Eq:HigherOrderCharacteristicFrequencies}
&\omega_p \approx \pm \sqrt N \gamma\sbkt{\frac{1}{\sqrt 2}-\frac{N\eta}{2\sqrt 2}+\frac{3}{4}\bkt{N\eta}^2+\mathcal{O}\sbkt{(N\eta)^3}}.
}
 The peaks closest to $\omega=\omega_0$ are split by $\approx\sqrt \frac{2N}{1+2N\eta}\gamma$ for $ N \eta \ll 1$ ~\cite{Guimond2016}. This splitting represents the characteristic coupling strength  of the emitter to its environment.
When the collective decay rate $N\gamma$ is significant such that $N\eta\sim1$, even if the delay is small such that $ \eta\ll1$,  we must include higher order terms in $N\eta$ to determine the splitting accurately. Thus the effects of delay and multimode dynamics can become significant as the cooperative linewidth of  emitter becomes comparable to the cavity FSR.  Fig.\,\ref{Fig:FrequencyResponseEtaRange} shows the interplay between the collectively enhanced emitter-environment coupling (due to $N$) and the delay effects (due to $ \eta$) that in turn act to reduce the effective cooperative coupling. In particular,  we observe that the effective emitter-environment coupling  increases with
$N$ as $\sim \sqrt N$ until $N \eta \sim 1$, where it saturates due to delay
effects. This illustrates limitations to cooperative atom-field coupling in the presence of time-delayed feedback.

\begin{figure*}[ht]
   \includegraphics[width = 1\linewidth]{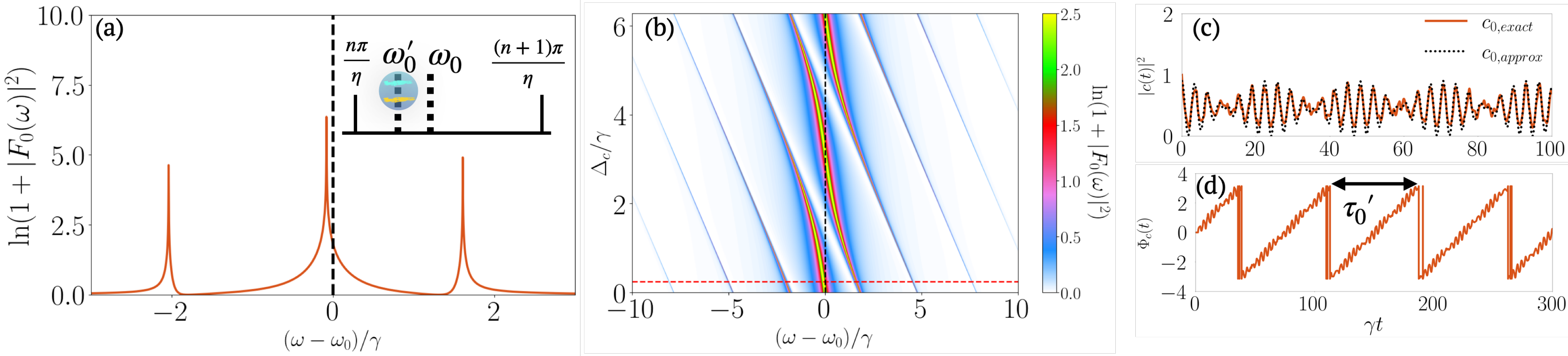}
    \caption{  Spectral density and atomic dynamics for $\eta\approx1$ with $\delta=0$, N=100 atoms in the arrays.  (a) For $\Delta_c=0.25\gamma$ between $\omega_0$ and the nearest cavity resonance, the spectral density shows the shifted emitter's resonance $\omega_0^\prime$ (central peak) and the two nearest frequencies $\omega_{\pm1}$ which manifest antinode at the emitter's position.   Schematic shows that the emitter is slightly displaced from cavity resonances exhibiting node at its position. (b) Semi-log plot of the spectral density as a function of increasing $\Delta_c$ such that $\Delta_c=0$ corresponds to a node at the emitter's location. The dashed-red horizontal line corresponds to $\Delta_c=0.25\gamma$.  (c) Emitter's dynamics (solid red) with $\Delta_c=0.25\gamma$ overlaid with the expected beat structure (black dotted). (d) Dynamical phase of the emitter as a function of time . }
    \label{Fig:NMFreqPull}
\end{figure*}

\subsection{Emitter at node}
If the emitter is placed at a node, a bound emitter-photon state is established in both the single-mode and multimode regimes (Fig.\,\ref{Fig:NodeNoDetuningCavity}(a),(b)), while the atomic mirrors remain unexcited. The $c_{0,approx}$ curve in Fig.\,\ref{Fig:NodeNoDetuningCavity}(a) shows the approximate emitter dynamics, obtained using the emitter's resonant frequency at $\omega=0$ and three characteristic frequencies each to the left and right of it, in good agreement with its exact numerical dynamics. In Fig.\,\ref{Fig:NodeNoDetuningCavity}(a) we see the time-domain multi-mode dynamics of the emitter, wherein the excitation probability does not go to zero at any point. This is because, when the emitter is placed at a node,  the reflected field destructively interferes with the emitter to inhibit its decay at time $2d/v$, at which point its dynamics deviates from free-space spontaneous decay (green-dashed curve). Fig.\,\ref{Fig:NodeNoDetuningCavity}(c) and (d) demonstrate this atom-photon bound  state   remain indefinitely excited inside the atomic cavity~\cite{Calajo, Sinha20a, Guo2020, Trivedi2021}.

We note from Fig.\,\ref{Fig:NodeNoDetuningCavity}(e) that in the single-mode limit the emitter's response is dominated by a single frequency at its resonance with a narrow linewidth implying the effective decay rate for the emitter is $\approx0$. As one goes towards larger  $\eta$ values, entering the non-Markovian regime, more frequencies contribute to the emitter's dynamics. The two peaks to the left and right (labeled $\omega_{-1}$ and $\omega_{+1}$ respectively) of the emitter's resonant frequency are separated by $\pi/(2\eta)$ from $\omega_0$. These correspond to cavity modes that exhibit antinodes at the emitter's position. Other adjacent cavity modes $\omega_{p\neq\pm 1}$ are separated by $\pi/\eta$, as dictated by the FSR of the half-cavity.  A comparison of Fig.\ref{Fig:AntiNodeNoDetuningCavity}(e) and \ref{Fig:NodeNoDetuningCavity}(e) demonstrates the siginifcant modification of the effective spectral density caused by the position of the emitter.

\section{Dynamics in the presence of detuning}
So far, we have assumed that the emitter is resonant with one of the cavity modes, as well as the internal resonance frequency $ \omega_M$ of the mirror atoms. We now introduce two detunings: (i) between the emitter and nearest cavity resonance such that $\Delta_c \equiv \omega_{c,n}-\omega_0$ (where for $\Delta_c=0$, $\omega_0=\omega_{c,n}= n\pi v/2d$ corresponds to a cavity mode exhibiting a node at the emitter's position); and (ii) a finite  detuning $ \delta$ between the emitter's resonance and that of the mirror atoms. 
\subsection{Frequency pulling for $ \Delta_c \neq 0 $} 
 In the single-mode regime,  the introduced propagation phase $\Delta_c~\frac{d}{v}\ll1$, considering that $ \Delta_c \lesssim \gamma$ and $ \gamma d/v\ll1$. Thus the shift in the cavity node location with respect to the emitter's position is negligible, leaving the system dynamics unaffected. When $\eta\approx1$, the cavity FSR is sufficiently small such that a small detuning $\Delta_c=0.25\gamma$ (chosen arbitrarily)  acts to `pull' the emitter's resonance frequency.  In Fig.\,\ref{Fig:NMFreqPull}(a) the spectral density shows that the cavity resonances shift relative to $\omega_0$ (dashed vertical line in  Fig.\,\ref{Fig:NMFreqPull}(a)) and pull the emitter's resonance to the new frequency $\omega_0^\prime$ (given by the central peak in Fig.\,\ref{Fig:NMFreqPull}(a)). This change in the emitter's resonance frequency, or frequency pulling,  can be computed by finding the pole nearest to $\omega=\omega_0$ in Eq.\,\eqref{Eq:DerivativeDenominator} for  $\phi_0=n\pi+\Delta_c\frac{d}{v}$.

 The cavity resonance peaks $\omega_{-1}$ and $\omega_{+1}$ have different relative separation from $\omega_0^\prime$ which results in the beat frequency $\abs{{\omega_{+1}+\omega_{-1}}}/{2}$  observed in Fig.\,\ref{Fig:NMFreqPull}(c). This is verified with the black-dotted curve plotted using Eq.\,\eqref{Eq:ApproximateEmitterDynamicsNM}.
 The emitter's excited state probability amplitude at any time $t$ can be expressed as $c_0(t) \equiv A(t) e^{ i  \Phi_c(t)}$ and we observe beats in the emitter's excitation probability and the dynamical phase $\Phi_c$.   For reference,  when $\Delta_c=0$, the phase $\Phi_c$ remains constant. We see from the frequency response of the emitter that the time-period  for the beat oscillations (zig-zag features in Fig.\,\ref{Fig:NMFreqPull}(d)) is given by $\tau_0^\prime=\frac{2\pi}{\abs{\omega_0^\prime}}$. 
 
 Fig.\,\ref{Fig:NMFreqPull}(b) shows that as we move up the $\Delta_c$ axis,   on an increment of $\Delta_c~\frac{d}{v}=\frac{\pi}{2}$, the emitter's resonance coincides with  an anti-node of the cavity and its dynamics mutates to that of an emitter placed at an antinode and the spectral density shows that the emitter's resonance is split into two peaks. On an increment of $\Delta_c~\frac{d}{v}=\pi$, the emitter's resonance coincides with  another node of the cavity. There is no frequency pulling on the emitter's resonance in both these cases and consequently, no beats in its dynamics.  Incrementing $\Delta_c$  cycles the emitter between nodes and antinodes in a long cavity. 
\subsection{Detuning between atomic resonances}
We now consider the dynamics of the system when there is a finite detuning between the emitter and the mirror atoms, with $\delta\neq0$. 
\begin{figure}[b]
   \includegraphics[width = 1\linewidth]{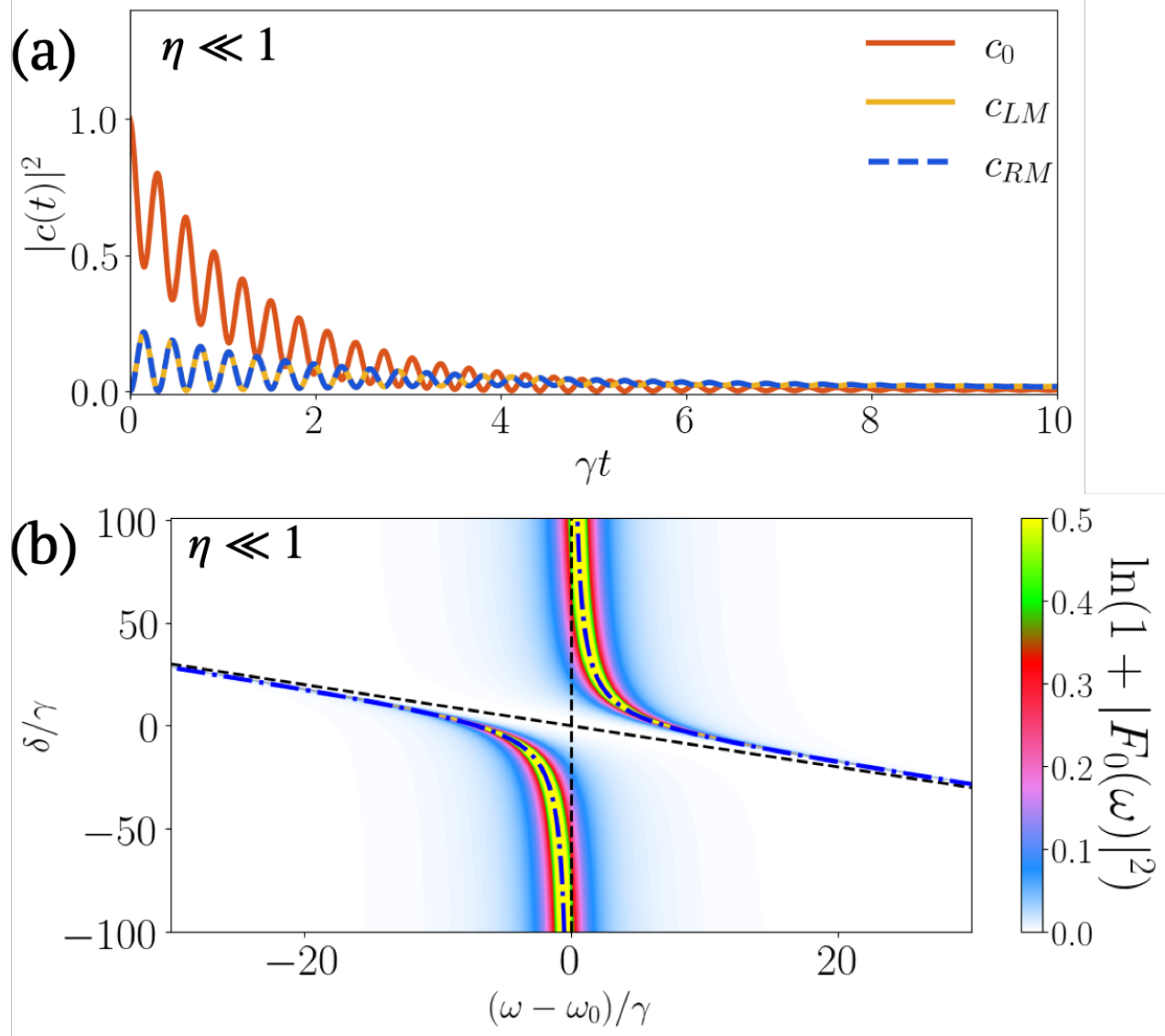}
    \caption[width = \linewidth]{  (a) Atomic dynamics based on Eq.\,\eqref{MarkovianCavity},\eqref{MarkovianCavityMirror} for internal detuning $\delta=15\gamma$ with $N=100$, $\phi_0=\pi/2$ and $N\eta\ll1$. (b) Emitter's frequency response $\abs{F_0(\omega)}^2$ as a function of $\delta$ and $\omega$. Blue-dashed-dotted curves are the energy eigenvalues (Eq.\,\eqref{Eq:AvoidedCrossingEnergyEigenvalues}) of the coupled emitter-array state and the black-dashed lines denote the asymptotic limits, $\omega_0$ and $\omega_M$, in case of large detuning. }
    \label{Fig:AvoidedCrossing}
\end{figure}
\begin{figure}[t]
   \includegraphics[width = 1\linewidth]{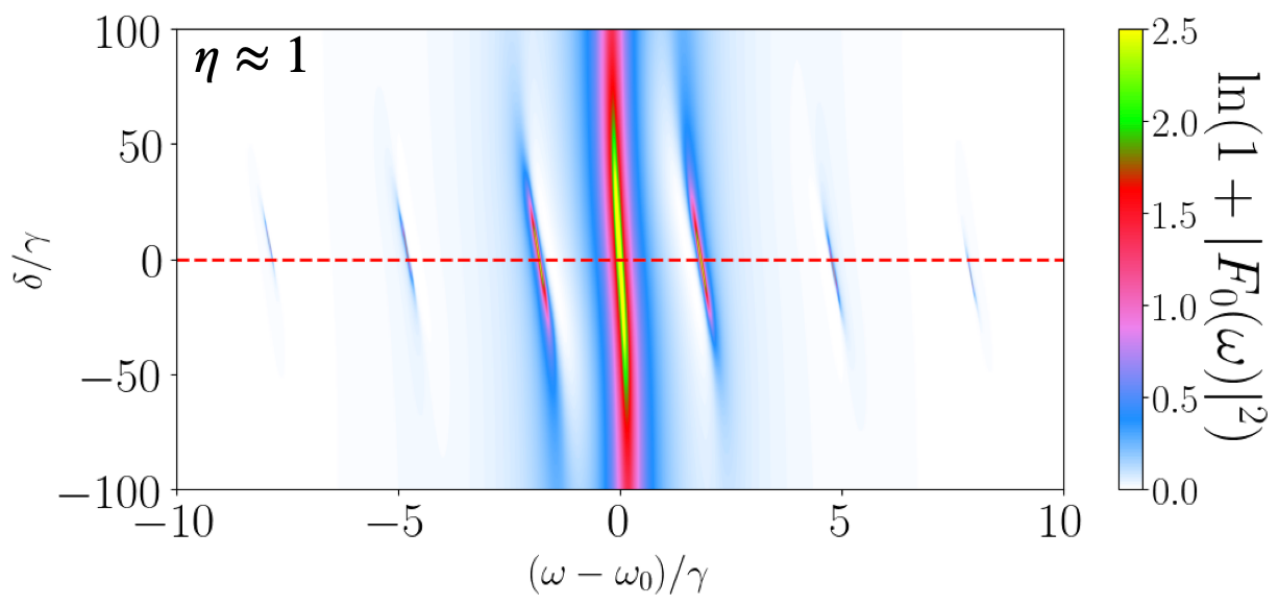}
    \caption[width = \linewidth]{ Emitter's frequency response   $\abs{F_0(\omega)}^2$ as a function of $\omega$ and the internal detuning $\delta$ for $N=100$, $\phi_0=n\pi$ and $\eta\approx1$}
    \label{Fig:InternalDetuning}
\end{figure}
In the single-mode regime, we arrive at analytical expressions for the emitter and atomic mirrors' dynamics  in the limit of large $N$ by solving Eqs.\,\eqref{CavityDDEa}-\eqref{CavityDDEc} under the following assumptions: $d/v$ is considered to be smaller than  the timescales associated with the collective coupling strength $(\sim 1/(N\gamma))$ and that for the emitter-mirror detuning $ 1/\delta $. Furthermore we assume the detuning to be much smaller than the collective coupling strength of the atomic mirrors, such that $d/v\ll 1/\delta \ll1/(N\gamma)$. When $\phi_0=\pi/2$, we obtain the emitter and the atomic array dynamics as
\begin{subequations}
    \begin{align}
        c_0(t) \approx &\frac{-i \delta-\frac{\gamma}{2}}{\sqrt{2N\gamma^2+\delta^2}}e^{-\Upsilon^\prime t/2}\sin\bkt{\sqrt{2N\gamma^2+\delta^2}\frac{ t}{2}}\non\\
        &+e^{-\Upsilon^\prime t/2}\cos\bkt{\sqrt{2N\gamma^2+\delta^2}\frac{ t}{2}}\label{MarkovianCavity} \\ 
        c_{LM,RM}(t) \approx &\sbkt{\frac{-\bkt{i \delta+\frac{\gamma}{2}}^2}{2\sqrt{2N\gamma^2+\delta^2}}-\frac{\sqrt{2N\gamma^2+\delta^2}}{2}}\non\\
        &\frac{e^{-\Upsilon^\prime t/2}e^{-i \delta t}e^{-i \phi_M}}{\sqrt N}\sin\bkt{\sqrt{2N\gamma^2+\delta^2}\frac{ t}{2}}\label{MarkovianCavityMirror}
    \end{align}
\end{subequations}
 where, $\Upsilon^\prime \equiv -{i \delta}+\frac{\gamma}{2}$, $\sqrt{2N\gamma^2+\delta^2}$ is identified as the effective Rabi frequency. 
 As expected from Fig.\,\ref{Fig:AntiNodeNoDetuningCavity}(a), the Rabi oscillation of the emitter is accompanied by an exponentially decaying envelope $e^{-\Upsilon^\prime t/2}$ which eventually leads to the atoms being completely de-excited. When the emitter and the atomic mirrors are perfectly resonant ($\delta=0$), the Rabi frequency is simply $\sqrt{2N}\gamma$ and  $\Upsilon\to\gamma/2$ in accordance with~\cite{Chang2012}.  
On introducing $\delta$, the Rabi frequency increases with $\delta$, and the  minima of the excitation probability in each cycle is no longer zero (Fig.\,\ref{Fig:AvoidedCrossing}(a)).  $\delta$ causes an avoided crossing in the frequency response where the eigenvalues 
\eqn{\label{Eq:AvoidedCrossingEnergyEigenvalues}
\omega = \pm\sqrt{2N\gamma^2+\delta^2}} resulting from the hybridization of the emitter and the array atoms' frequencies, given by the blue-dashed-dotted curves  in Fig.\,\ref{Fig:AvoidedCrossing}(b). The eigenvalues asymptotically approach the resonant frequencies of the emitter and the array atoms, with the emitter's response converging to a single Lorentzian with width $\gamma$.

As we increase $d$ to enter the multimode regime,  Fig.\,\ref{Fig:InternalDetuning} shows that increasing $\abs \delta$ diminishes the emitter's coupling to the cavity modes such that for large $\delta$, the emitter's dynamics is governed by a single Lorentzian frequency response at the emitter's resonance. Comparing  Fig.\,\ref{Fig:NMFreqPull}(b) and Fig.\,\ref{Fig:InternalDetuning}  shows that qualitatively the detuning $\delta$ differs from $\Delta_c$ in that increasing $\delta$ introduces loss to the atomic mirror cavity, attenuating the response at antinode frequencies away from the emitter's resonance, whereas changing $\Delta_c$ pulls the emitter's resonance frequency cyclically between a node and an antinode without making the cavity lossy.

 \section{Discussion}

We have analyzed the non-Markovian spontaneous emission dynamics of an excited two-level emitter placed between a pair of two-level atomic arrays forming a cavity. 
As the cavity round-trip time $ \sim d/v$ becomes comparable to the cooperatively broadened decay time of the atoms $ \sim 1/(N\gamma)$,  the system dynamics is governed by time-delayed feedback and multimode field effects.  We illustrate such dynamics for two specific configurations wherein the emitter is placed at an effective  node or an antinode of the atomic cavity (Fig.\,\ref{Fig:AntiNodeNoDetuningCavity} and \ref{Fig:NodeNoDetuningCavity}).  We show that such a system exhibits cooperative and multimode light-matter coupling that strongly modifies the effective spectral density of the field, as shown in Fig.\,\ref{Fig:AntiNodeNoDetuningCavity}\,(e) and \ref{Fig:NodeNoDetuningCavity}\,(e). Such modified spectral density facilitates efficient computation of the non-Markovian dynamics of the emitter in terms of a few prominent characteristic frequencies, as demonstrated in Fig.\,\ref{Fig:AntiNodeNoDetuningCavity}\,(a) and \ref{Fig:NodeNoDetuningCavity}\,(a).

Additionally, we characterize the  effective light-matter coupling  strength via the vacuum Rabi splitting. We study the effect of delay and number of atoms on this effective coupling strength, showing that it increases with $N$ as $ \sim \sqrt{N}$ until $ N \eta\sim1$ where  it saturates due to  delay effects  (Fig.\,\ref{Fig:FrequencyResponseEtaRange}).  


Adding a  detuning  $\Delta_c$ between the emitter and  cavity resonance, we observe frequency pulling of the emitter's resonance in the presence of multimode strong coupling (Fig.\,\ref{Fig:NMFreqPull}).  Finally, we analyze the dynamics of the emitter and the effective spectral density in the presence of a detuning between the emitter and the internal resonance of the atomic mirrors. 

Our results are pertinent to waveguide-QED setups that exhibit cooperative interactions between multiple emitters in the presence of time-delayed feedback, across various platforms, such as neutral atoms coupled to optical nanofibers~\cite{Corzo2016,Sorensen2016}, circuit QED~\cite{Mirhosseini2019} and quantum emitters coupled via atomic matter waves~\cite{Kim2025}. 
Our results demonstrate the limitations of collectively enhanced light-matter coupling in the presence of time-delayed feedback, motivating a careful consideration of delay when harnessing cooperative effects in waveguide QED.

We further illustrate that a spectral analysis of non-Markovian dynamics allows one to efficiently capture the system dynamics in terms of the few emergent modes of the bath~\cite{keefe2024, cilluffo2024}. Such a description is pertinent for keeping account of how energy and quantum information is distributed across different modes of the EM field in long-distance waveguide-coupled quantum emitters. This can enable strategic readout and manipulation of quantum information by monitoring and processing only a few relevant modes of the bath.



 \section{Acknowledgment}

 We are thankful to Archana Kamal, Chloe I. Marzano, Dominik Schneble, Hakan T\"ureci, Murray Holland and Sang-Eon Bak   for  helpful discussions.  
K.S. acknowledges support from the National Science Foundation under Grant No. PHY-2309341, by the John Templeton Foundation under Award No. 62422,  and Army Research Office under Award No. W911NF2410080. P.S. acknowledges support from ANID, through FONDECYT Grant No. 1240204. This research was supported in part by grant NSF PHY-2309135 to the Kavli Institute for Theoretical Physics (KITP).

\appendix 
\begin{widetext}
\section{Coupled atom-field dynamics}
\label{AppendixA}

We solve the Schr\"odinger equation using the  interaction Hamiltonian in the interaction picture (Eq.\eqref{CavityHamiltonian}), for the state ansatz in Eq.\,\eqref{CavityAnsatz} in the main text to obtain the equations of motion for the atomic excitation amplitudes as follows:

    \eqn{\label{Eq:EmitterInTermsofField}
\der{c_0 (t)}{t} = &-i   \int_{-\infty}^\infty dk g_k c_{ k} (t) e^{-i (\omega - \omega_0 )t }\\
\label{Eq:LeftMirrorInTermsofField}
\der{c_{e_{n}, G'} (t) }{t} = &-i   \int_{-\infty}^\infty dk g_k c_{ k}(t)e^{i k x_n } e^{-i (\omega - \omega_M )t }\\
\label{Eq:RightMirrorInTermsofField}
\der{c_{ G',e_{n}} (t) }{t} = &-i   \int_{-\infty}^\infty dk g_k c_{ k}(t)e^{i k x_n }  e^{-i (\omega - \omega_M )t }
}

   Similarly for the field excitation amplitudes:
\eqn{
\der{c_{ k}(t)}{t} = & -ig^*_k\sbkt{  c_0 (t)e^{i (\omega - \omega_0 )t}+\cbkt{\sum_{n=-N}^{-1}  c_{e_n,G'} (t) e^{-i k x_n } + \sum_{n=1}^N  c_{G',e_n} (t) e^{-i k x_n }}e^{i (\omega - \omega_M )t}} \label{Eq:FieldEquationsofMotion}
}

We formally integrate Eq.\,\eqref{Eq:FieldEquationsofMotion} of motion and substitute in Eqs.\eqref{Eq:EmitterInTermsofField}--\eqref{Eq:RightMirrorInTermsofField} to  obtain Eqs. \eqref{CavityDDEa}--\eqref{CavityDDEc} in the main text.

The Laplace transforms of the coupled delay differential equations for the atomic excitation amplitudes (Eqs. \eqref{CavityDDEa}--\eqref{CavityDDEc}) are:

\begin{subequations}
    \begin{align}
        \bkt{s+\frac{\gamma}{2}}\tilde{c}_0(s) = c_0(0) - \frac{\gamma\sqrt{N}}{2}\bkt{\tilde{c}_{LM}(s-i\delta)e^{-sd/v}e^{i\omega_0d/v}+\tilde{c}_{RM}(s-i\delta)e^{-sd/v}e^{i\omega_0d/v}}\label{LaplaceCavityEmitter} \\
        \bkt{s+\frac{\gamma N}{2}}\tilde{c}_{LM}(s) = c_{LM}(0) - \frac{\gamma\sqrt{N}}{2}\tilde{c}_0(s+i\delta)e^{-sd/v}e^{i\omega_Md/v}-\frac{\gamma N}{2}\tilde{c}_{RM}(s)e^{-2sd/v}e^{2i\omega_Md/v}\label{LaplaceCavityLM}  \\
        \bkt{s+\frac{\gamma N}{2}}\tilde{c}_{RM}(s) = c_{RM}(0) - \frac{\gamma\sqrt{N}}{2}\tilde{c}_0(s+i\delta)e^{-sd/v}e^{i\omega_Md/v}-\frac{\gamma N}{2}\tilde{c}_{LM}(s)e^{-2sd/v}e^{2i\omega_Md/v} \label{LaplaceCavityRM}
    \end{align}
\end{subequations}
which can be decoupled and solved numerically as we discuss below.

\section{Response Functions}
\label{Appendix:ResponseFunctions}
Decoupling the set of equations Eq.\eqref{LaplaceCavityEmitter} - \eqref{LaplaceCavityRM} we have:

    \eqn{
    \label{Eq:MirrorLaplaceInv}
    \tilde{c}_0(s)= &\frac{e^{2
     i   \phi_0}N\gamma+e^{2sd/v}\bkt{N\gamma+2s-2 i  \delta}}{\bkt{\frac{\gamma}{2}+s}\bkt{e^{2 i  \phi_0}N\gamma +e^{2sd/v}\bkt{N\gamma+2s-2 i  \delta}}-N\gamma^2e^{2 i  \phi_0}}\non\\
    \tilde{c}_{LM}(s)=& \frac{-2 \gamma\sqrt N e^{sd/v}e^{
     i   \phi_0}}{-2e^{2 i  \phi_0- i  \delta d/v}N\gamma^2+e^{2 i  \phi_0- i  \delta d/v}N\gamma\bkt{\gamma+2s+2 i  \delta}+e^{\bkt{2s+ i  \delta} d/v}(N\gamma+2s)(\gamma+2s+2 i  \delta)}
    =\tilde{c}_{RM}(s).
    }
    
Changing variables from s$\to-i \omega$ gives us the  response of the emitter and the atomic mirrors for real frequencies:

       \eqn{
     &{c}_0(t)= \frac{1}{2\pi i }\int_{-\infty}^{ \infty}d\omega\frac{e^{2
    i\phi_0}N\gamma+e^{-2 i  \omega\eta}\bkt{N-2 i \omega-2i\delta}}{\bkt{\frac{\gamma}{2}- i  \omega}\bkt{e^{2 i  \phi_0}N\gamma +e^{-2i \omega d/v}\bkt{N\gamma-2 i  \omega-2i\delta}}-N\gamma^2e^{2 i  \phi_0}}e^{- i  \omega t}=\int_{-\infty}^{\infty}d\omega F_0(\omega)e^{- i  \omega t}\non\\
    &{c}_{LM}(t)= {c}_{RM}(t)=\int_{-\infty}^{\infty}d\omega F_M(\omega)e^{- i  \omega t}=\non \\ 
   & \frac{1}{2\pi i }\int_{- \infty}^{ \infty}d\omega\frac{-2 \gamma\sqrt N e^{- i  \omega d/v}e^{     i   \phi_0}e^{- i  \omega t}}{-2e^{2 i  \phi_0- i  \delta d/v}N\gamma^2+e^{2 i  \phi_0- i  \delta d/v}N\gamma\bkt{\gamma-2 i  \omega+ 2i  \delta}+e^{\bkt{2s+i\delta}d/v}(N\gamma-2 i  \omega)(\gamma-2 i  \omega+2i  \delta)}
    }
where  $F_0(\omega)$ is given by Eq.\,\eqref{EmitterResponseFunction}. 
The poles gives of the above integrands correspond to the characteristic frequencies present in the atomic dynamics.
We find the poles numerically using Muller's method. For $\eta<1$, we Taylor-expand the denominator about $\eta=0$ giving us 
\eqn{\label{Eq:TaylorResponseDenominatorAntiNode} 
D_0\bkt{\omega}=&\gamma^2\bkt{\delta+\omega}^2+\bkt{-N\gamma^2+2\omega\bkt{\delta+\omega}}^2-2\eta\bkt{N\gamma^3\delta\omega+N\gamma^3\omega^2+2N^2\gamma^3\omega^2-4N\gamma\delta\omega^3-4N\gamma\omega^4}\non\\
&+\eta^2\left(-N^2\gamma^4\omega^2+8N\gamma^2\delta\omega^3+8n\gamma^3\omega^4+4N^2\gamma^2\omega^4\right)+\frac{4}{3}\eta^3\omega^3\bkt{N\gamma^3\delta+N\gamma^3\omega+2N^2\gamma^3\omega-4N\gamma\delta\omega^2\right.\non\\ 
&\left.-4N\gamma\omega^3}+\mathcal{O}\sbkt{\eta}^4  
}
for the specific case of $\phi_0=(2n+1)\pi/2$ and
\eqn{\label{Eq:TaylorResponseDenominatorNode}
D_0\bkt{\omega}=4\omega^2\bkt{\delta+\omega}^2+4\gamma^2\bkt{\delta+\omega+2N\omega}^2+2\eta\bkt{N\gamma^3\delta\omega+N\gamma^3\omega^2+2N^2\gamma^3\omega^2-4N\gamma\delta\omega^3-4N\gamma\omega^4}+\non\\
\eta^2\omega^2\gamma^2\bkt{N^2\gamma^2-8N\delta\omega-8N\omega^2-4N\omega^2}-\frac{4}{3}\eta^3\omega^3\bkt{N\gamma^3\delta+N\gamma^3\omega+2N^2\gamma^3\omega-4N\gamma\delta\omega^2-4N\gamma\omega^4}+\mathcal{O}[\eta]^4
}
for $\phi_0=n\pi$. Analytical expression for the effective vacuum Rabi splitting for an emitter (Eq.\,\eqref{Eq:HigherOrderCharacteristicFrequencies}) at antinode is obtained to $\mathcal{O}\sbkt{\eta}^3$ accuracy by finding the roots  of Eq.\,\eqref{Eq:TaylorResponseDenominatorAntiNode} to second order in $ \eta$.
\section{Intensity}
\label{AppendixIntensity}
The field intensity can be written by splitting up the electric field E operator given by $\hat{E}\bkt{x,t} = \int_{-\infty}^\infty dk {E}_k \hat{a}_ke^{i  k x}e^{- i \omega t}$~~ as given in the main text  into its right going \textit{a} modes and left going \textit{b} modes as $\hat{E}\bkt{x,t} = \int_0^\infty d\omega {E}_\omega \bkt{\hat{a}_\omega e^{i  \omega x/v}+\hat{b}_\omega e^{-i  \omega x/v}}e^{- i \omega t}$  
, with $E_\omega = E_k/\sqrt v$ and $a_\omega = a_k/\sqrt v$. Considering $E_\omega$ to be constant around $\omega_0$, the intensity is:

\eqn{\label{A8}
I (x,t) = &{\epsilon_0v}\abs{E_{\omega_0}}^2\bra{\Psi (t) }\int_0^\infty d\omega' \sbkt{\hat a_{\omega'}^\dagger e^{-i  \omega' x/v} + \hat b_{\omega'}^\dagger e^{i  \omega' x/v} } e^{i  \omega' t} \int_0^\infty d\omega \sbkt{\hat a_{\omega} e^{- i  \omega x/v} + \hat b_{\omega} e^{i  \omega x/v} }e^{-i  \omega t}\ket{\Psi (t) }\non \\
&= {\epsilon_0v}\abs{E_{\omega_0}}^2\int_0^\infty d\omega' \int_0^\infty d\omega \bra{G}\bra{g}\cbkt{(c_a^*(\omega',t)) e^{-i  \omega' x/v} + (c_b^*(\omega',t)) e^{i \omega' x/v}} \non\\
 &\cbkt{ c_a(\omega,t) e^{i \omega x/v} + c_b(\omega,t) e^{-i \omega x/v} } \ket{G}\ket{g} e^{i (\omega' - \omega) t}\\ 
&={\epsilon_0v}\abs{E_{\omega_0}}^2\abs{\int_0^\infty d\omega \sbkt{c_a(\omega,t) e^{i \omega x/v} + c_b(\omega,t) e^{-i  \omega x/v}  }e^{-i  \omega t}}^2
}

Here we can substitute the equations of motion for the field Eq.\,\eqref{Eq:FieldEquationsofMotion} in terms of the atomic dynamics. 

\eqn{\label{Eq:IntensityUnintegrated}
 I (x,t) &=\frac{\epsilon_0 v\gamma}{2\pi}\abs{E_{\omega_0}}^2\left|\int_0^\infty d\omega e^{-i \omega t}\int_0^t d\tau\left[c_0(\tau)\bkt{e^{i \omega\frac{x}{v}}+e^{-i \omega\frac{x}{v}}}e^{i (\omega-\omega_0)\tau}\right.\right.\non\\
 &\left.\left.+\sqrt{N}c_{RM}(\tau)\bkt{e^{i \omega\frac{x-d}{v}}+e^{-i \omega\frac{x-d}{v}}}e^{i (\omega-\omega_M)\tau}+\sqrt{N}c_{LM}(\tau)\bkt{e^{i \omega\frac{x+d}{v}}+e^{-i \omega\frac{x+d}{v}}}e^{i (\omega-\omega_M)\tau}\right]\right|^2\non \\
\implies \frac{I(x,t)}{I_0} &=\left|\int_0^t d\tau\left[c_0(\tau)\bkt{\delta\bkt{\tau-t+\frac{x}{v}}+\delta\bkt{\tau-t-\frac{x}{v}}}e^{-i \omega_0\tau}\right.\right.\non\\
&\left.\left.+\sqrt{N}c_{RM}(\tau)\bkt{\delta\bkt{\tau-t+\frac{x-d}{v}}+\delta\bkt{\tau-t-\frac{x-d}{v}}}e^{-i \omega_M\tau}\right.\right.\non\\
&\left.\left.+\sqrt{N}c_{LM}(\tau)\bkt{\delta\bkt{\tau-t+\frac{x+d}{v}}+\delta\bkt{\tau-t-\frac{x+d}{v}}}e^{-i \omega_M\tau}\right]\right|^2
}



 with $I_0$  as defined in the main text. Performing the integral in Eq.\,\eqref{Eq:IntensityUnintegrated} gives us Eq.\,\eqref{Eq:FieldIntensity}.

\section{Rabi Oscillations in single-mode regime}
 In the single-mode regime, the decoupled Laplace transformations for Eqs.~\eqref{LaplaceCavityEmitter}-\eqref{LaplaceCavityRM} take the form

\eqn{\label{EmitterDecoupledLaplace}
        \tilde{c}_0(s)=&\frac{2}{\gamma+2s-\frac{e^{i \phi_0}N\gamma^2}{\bkt{N\gamma+s-i \delta}\cos\bkt{\phi_M}-\bkt{i  s+\delta}\sin\bkt{\phi_M}}}
        \\
    \label{MirrorrDecoupledLaplace}
        \tilde{c}_{LM}(s)=&-\frac{e^{i \phi_M}\gamma\sqrt N}{s\gamma+Ns\gamma+2s^2+i  \delta N\gamma+2i \delta s+ e^{2i \phi_M}N(s+i \delta)\gamma+e^{i \phi_M}N\gamma^2(-i +\cos\bkt{\phi_M})}=\tilde{c}_{RM}(s)
    }

When the emitter is at an antinode, its Laplace transform Eq.\,\eqref{EmitterDecoupledLaplace} is simplified by setting $\phi_0=\pi/2$, $\phi_M=\phi_0-\delta d/v$ and written as 

\eqn{
\tilde{c}_0(s) = \frac{2N\gamma \sin(\delta d/v)+2(s-i \delta)(\sin(\delta d/v)-i  \cos(\delta d/v))}{(\gamma+2s)(N\gamma+s-i \delta)\sin(\delta d/v) -i (\gamma+2s)(s-i \delta)\cos(\delta d/v)-i  N\gamma^2}
}

Inverting the Laplace transform for the emitter involves evaluating the poles of the quadratic polynomial denominator in s. We simplify this equation further by writing

\eqn{
\tilde{c}_0(s) = \frac{i  N\gamma \sin(\delta d/v) e^{-i \delta d/v}-i \delta+s}{s^2 +s\underbrace{\bkt{-i \delta+\frac{\gamma}{2}+i  N \gamma \sin(\delta d/v) e^{-i \delta d/v}}}_\Upsilon+\underbrace{\bkt{\frac{i  \gamma N}{2}\sin(\delta d/v) e^{-i \delta d/v}-\frac{i \delta}{2}+\frac{N\gamma}{2}e^{-i \delta d/v}}}_{\zeta}}
}
such that the roots of the denominator are $s_\pm=\frac{1}{2}\bkt{-\Upsilon\pm\sqrt{\Upsilon^2-4\zeta}}$. Using the Heaviside theorem~\cite{stegun}, we can invert this equation from  Laplace to time domain:
\eqn{
c_0(t) \approx \sum_{s=s_\pm}\frac{i  N \sin(\delta\eta) e^{-i \delta\eta}-i \delta+s}{2s+\Upsilon}e^{st}.
}
 In the limit where $N\to\infty$ such that $\delta/N\gamma\ll 1$, $\delta d/v\ll 1$ and $N\gamma d/v\ll 1$, we can simplify the expression $\sqrt{\Upsilon^2-4\zeta}\to\sqrt{2N\gamma^2+\delta^2}$ to get Eq.\,\eqref{MarkovianCavity}. We infer the equality of left and right arrays' dynamics from Eq.\,\eqref{MirrorrDecoupledLaplace}. Instead of inverting the Laplace transforms for $\tilde{c}_{LM}(s)$ and $\tilde{c}_{RM}(s)$, we substitute Eq.\,\eqref{MarkovianCavity} into Eq.\,\eqref{CavityDDEa} by considering the retardation to be negligible ($\gamma d/v\ll1$) to obtain Eq.\,\eqref{MarkovianCavityMirror}. Under the same assumptions, setting $\phi_0 =\pi$ yields the atomic dynamics when the emitter is placed at a node with $c_0(t)\approx1$ and $c_{LM}(t)=c_{RM}(t)\approx0$ with $\delta=0$.

\end{widetext}

\bibliography{AtomicMirror.bib}

\begin{thebibliography}{45}%
\makeatletter
\providecommand \@ifxundefined [1]{%
 \@ifx{#1\undefined}
}%
\providecommand \@ifnum [1]{%
 \ifnum #1\expandafter \@firstoftwo
 \else \expandafter \@secondoftwo
 \fi
}%
\providecommand \@ifx [1]{%
 \ifx #1\expandafter \@firstoftwo
 \else \expandafter \@secondoftwo
 \fi
}%
\providecommand \natexlab [1]{#1}%
\providecommand \enquote  [1]{``#1''}%
\providecommand \bibnamefont  [1]{#1}%
\providecommand \bibfnamefont [1]{#1}%
\providecommand \citenamefont [1]{#1}%
\providecommand \href@noop [0]{\@secondoftwo}%
\providecommand \href [0]{\begingroup \@sanitize@url \@href}%
\providecommand \@href[1]{\@@startlink{#1}\@@href}%
\providecommand \@@href[1]{\endgroup#1\@@endlink}%
\providecommand \@sanitize@url [0]{\catcode `\\12\catcode `\$12\catcode `\&12\catcode `\#12\catcode `\^12\catcode `\_12\catcode `\%12\relax}%
\providecommand \@@startlink[1]{}%
\providecommand \@@endlink[0]{}%
\providecommand \url  [0]{\begingroup\@sanitize@url \@url }%
\providecommand \@url [1]{\endgroup\@href {#1}{\urlprefix }}%
\providecommand \urlprefix  [0]{URL }%
\providecommand \Eprint [0]{\href }%
\providecommand \doibase [0]{http://dx.doi.org/}%
\providecommand \selectlanguage [0]{\@gobble}%
\providecommand \bibinfo  [0]{\@secondoftwo}%
\providecommand \bibfield  [0]{\@secondoftwo}%
\providecommand \translation [1]{[#1]}%
\providecommand \BibitemOpen [0]{}%
\providecommand \bibitemStop [0]{}%
\providecommand \bibitemNoStop [0]{.\EOS\space}%
\providecommand \EOS [0]{\spacefactor3000\relax}%
\providecommand \BibitemShut  [1]{\csname bibitem#1\endcsname}%
\let\auto@bib@innerbib\@empty
\bibitem [{\citenamefont {Dicke}(1954)}]{Dicke1954}%
  \BibitemOpen
  \bibfield  {author} {\bibinfo {author} {\bibfnamefont {R.~H.}\ \bibnamefont {Dicke}},\ }\href {\doibase 10.1103/PhysRev.93.99} {\bibfield  {journal} {\bibinfo  {journal} {Phys. Rev.}\ }\textbf {\bibinfo {volume} {93}},\ \bibinfo {pages} {99} (\bibinfo {year} {1954})}\BibitemShut {NoStop}%
\bibitem [{\citenamefont {Gross}\ and\ \citenamefont {Haroche}(1982)}]{Gross1982}%
  \BibitemOpen
  \bibfield  {author} {\bibinfo {author} {\bibfnamefont {M.}~\bibnamefont {Gross}}\ and\ \bibinfo {author} {\bibfnamefont {S.}~\bibnamefont {Haroche}},\ }\href {\doibase https://doi.org/10.1016/0370-1573(82)90102-8} {\bibfield  {journal} {\bibinfo  {journal} {Physics Reports}\ }\textbf {\bibinfo {volume} {93}},\ \bibinfo {pages} {301} (\bibinfo {year} {1982})}\BibitemShut {NoStop}%
\bibitem [{\citenamefont {Guerin}\ \emph {et~al.}(2017)\citenamefont {Guerin}, \citenamefont {Rouabah},\ and\ \citenamefont {Kaiser}}]{Guerin2017}%
  \BibitemOpen
  \bibfield  {author} {\bibinfo {author} {\bibfnamefont {W.}~\bibnamefont {Guerin}}, \bibinfo {author} {\bibfnamefont {M.}~\bibnamefont {Rouabah}}, \ and\ \bibinfo {author} {\bibfnamefont {R.}~\bibnamefont {Kaiser}},\ }\href {\doibase 10.1080/09500340.2016.1215564} {\bibfield  {journal} {\bibinfo  {journal} {Journal of modern optics}\ }\textbf {\bibinfo {volume} {64}},\ \bibinfo {pages} {895} (\bibinfo {year} {2017})}\BibitemShut {NoStop}%
\bibitem [{\citenamefont {Hammerer}\ \emph {et~al.}(2010)\citenamefont {Hammerer}, \citenamefont {S\o{}rensen},\ and\ \citenamefont {Polzik}}]{Hammerer2010}%
  \BibitemOpen
  \bibfield  {author} {\bibinfo {author} {\bibfnamefont {K.}~\bibnamefont {Hammerer}}, \bibinfo {author} {\bibfnamefont {A.~S.}\ \bibnamefont {S\o{}rensen}}, \ and\ \bibinfo {author} {\bibfnamefont {E.~S.}\ \bibnamefont {Polzik}},\ }\href {\doibase 10.1103/RevModPhys.82.1041} {\bibfield  {journal} {\bibinfo  {journal} {Rev. Mod. Phys.}\ }\textbf {\bibinfo {volume} {82}},\ \bibinfo {pages} {1041} (\bibinfo {year} {2010})}\BibitemShut {NoStop}%
\bibitem [{\citenamefont {Wang}\ \emph {et~al.}(2007)\citenamefont {Wang}, \citenamefont {Yelin}, \citenamefont {C\^ot\'e}, \citenamefont {Eyler}, \citenamefont {Farooqi}, \citenamefont {Gould}, \citenamefont {Ko\ifmmode~\check{s}\else \v{s}\fi{}trun}, \citenamefont {Tong},\ and\ \citenamefont {Vrinceanu}}]{Wang2007}%
  \BibitemOpen
  \bibfield  {author} {\bibinfo {author} {\bibfnamefont {T.}~\bibnamefont {Wang}}, \bibinfo {author} {\bibfnamefont {S.~F.}\ \bibnamefont {Yelin}}, \bibinfo {author} {\bibfnamefont {R.}~\bibnamefont {C\^ot\'e}}, \bibinfo {author} {\bibfnamefont {E.~E.}\ \bibnamefont {Eyler}}, \bibinfo {author} {\bibfnamefont {S.~M.}\ \bibnamefont {Farooqi}}, \bibinfo {author} {\bibfnamefont {P.~L.}\ \bibnamefont {Gould}}, \bibinfo {author} {\bibfnamefont {M.}~\bibnamefont {Ko\ifmmode~\check{s}\else \v{s}\fi{}trun}}, \bibinfo {author} {\bibfnamefont {D.}~\bibnamefont {Tong}}, \ and\ \bibinfo {author} {\bibfnamefont {D.}~\bibnamefont {Vrinceanu}},\ }\href {\doibase 10.1103/PhysRevA.75.033802} {\bibfield  {journal} {\bibinfo  {journal} {Phys. Rev. A}\ }\textbf {\bibinfo {volume} {75}},\ \bibinfo {pages} {033802} (\bibinfo {year} {2007})}\BibitemShut {NoStop}%
\bibitem [{\citenamefont {Goban}\ \emph {et~al.}(2015)\citenamefont {Goban}, \citenamefont {Hung}, \citenamefont {Hood}, \citenamefont {Yu}, \citenamefont {Muniz}, \citenamefont {Painter},\ and\ \citenamefont {Kimble}}]{Goban2015}%
  \BibitemOpen
  \bibfield  {author} {\bibinfo {author} {\bibfnamefont {A.}~\bibnamefont {Goban}}, \bibinfo {author} {\bibfnamefont {C.-L.}\ \bibnamefont {Hung}}, \bibinfo {author} {\bibfnamefont {J.~D.}\ \bibnamefont {Hood}}, \bibinfo {author} {\bibfnamefont {S.-P.}\ \bibnamefont {Yu}}, \bibinfo {author} {\bibfnamefont {J.~A.}\ \bibnamefont {Muniz}}, \bibinfo {author} {\bibfnamefont {O.}~\bibnamefont {Painter}}, \ and\ \bibinfo {author} {\bibfnamefont {H.~J.}\ \bibnamefont {Kimble}},\ }\href {\doibase 10.1103/PhysRevLett.115.063601} {\bibfield  {journal} {\bibinfo  {journal} {Phys. Rev. Lett.}\ }\textbf {\bibinfo {volume} {115}},\ \bibinfo {pages} {063601} (\bibinfo {year} {2015})}\BibitemShut {NoStop}%
\bibitem [{\citenamefont {Chen}\ \emph {et~al.}(2010)\citenamefont {Chen}, \citenamefont {Zhang}, \citenamefont {Xiao}, \citenamefont {Liang},\ and\ \citenamefont {Jia}}]{Chen2010}%
  \BibitemOpen
  \bibfield  {author} {\bibinfo {author} {\bibfnamefont {G.}~\bibnamefont {Chen}}, \bibinfo {author} {\bibfnamefont {Y.}~\bibnamefont {Zhang}}, \bibinfo {author} {\bibfnamefont {L.}~\bibnamefont {Xiao}}, \bibinfo {author} {\bibfnamefont {J.-Q.}\ \bibnamefont {Liang}}, \ and\ \bibinfo {author} {\bibfnamefont {S.}~\bibnamefont {Jia}},\ }\href {\doibase 10.1364/OE.18.023016} {\bibfield  {journal} {\bibinfo  {journal} {Opt. Express}\ }\textbf {\bibinfo {volume} {18}},\ \bibinfo {pages} {23016} (\bibinfo {year} {2010})}\BibitemShut {NoStop}%
\bibitem [{\citenamefont {Chang}\ \emph {et~al.}(2012)\citenamefont {Chang}, \citenamefont {Jiang}, \citenamefont {Gorshkov},\ and\ \citenamefont {Kimble}}]{Chang2012}%
  \BibitemOpen
  \bibfield  {author} {\bibinfo {author} {\bibfnamefont {D.~E.}\ \bibnamefont {Chang}}, \bibinfo {author} {\bibfnamefont {L.}~\bibnamefont {Jiang}}, \bibinfo {author} {\bibfnamefont {A.~V.}\ \bibnamefont {Gorshkov}}, \ and\ \bibinfo {author} {\bibfnamefont {H.~J.}\ \bibnamefont {Kimble}},\ }\href {\doibase 10.1088/1367-2630/14/6/063003} {\bibfield  {journal} {\bibinfo  {journal} {New Journal of Physics}\ }\textbf {\bibinfo {volume} {14}},\ \bibinfo {pages} {063003} (\bibinfo {year} {2012})}\BibitemShut {NoStop}%
\bibitem [{\citenamefont {Ruddell}\ \emph {et~al.}(2017)\citenamefont {Ruddell}, \citenamefont {Webb}, \citenamefont {Herrera}, \citenamefont {Parkins},\ and\ \citenamefont {Hoogerland}}]{Ruddell2017}%
  \BibitemOpen
  \bibfield  {author} {\bibinfo {author} {\bibfnamefont {S.~K.}\ \bibnamefont {Ruddell}}, \bibinfo {author} {\bibfnamefont {K.~E.}\ \bibnamefont {Webb}}, \bibinfo {author} {\bibfnamefont {I.}~\bibnamefont {Herrera}}, \bibinfo {author} {\bibfnamefont {A.~S.}\ \bibnamefont {Parkins}}, \ and\ \bibinfo {author} {\bibfnamefont {M.~D.}\ \bibnamefont {Hoogerland}},\ }\href {\doibase 10.1364/OPTICA.4.000576} {\bibfield  {journal} {\bibinfo  {journal} {Optica}\ }\textbf {\bibinfo {volume} {4}},\ \bibinfo {pages} {576} (\bibinfo {year} {2017})}\BibitemShut {NoStop}%
\bibitem [{\citenamefont {Li}\ \emph {et~al.}(2023)\citenamefont {Li}, \citenamefont {Pan}, \citenamefont {Liu}, \citenamefont {Zhou}, \citenamefont {Huang}, \citenamefont {Shen}, \citenamefont {Wang}, \citenamefont {Li},\ and\ \citenamefont {Guo}}]{Li2023}%
  \BibitemOpen
  \bibfield  {author} {\bibinfo {author} {\bibfnamefont {L.}~\bibnamefont {Li}}, \bibinfo {author} {\bibfnamefont {Y.-H.}\ \bibnamefont {Pan}}, \bibinfo {author} {\bibfnamefont {Y.-J.}\ \bibnamefont {Liu}}, \bibinfo {author} {\bibfnamefont {X.-L.}\ \bibnamefont {Zhou}}, \bibinfo {author} {\bibfnamefont {D.-Y.}\ \bibnamefont {Huang}}, \bibinfo {author} {\bibfnamefont {Z.-M.}\ \bibnamefont {Shen}}, \bibinfo {author} {\bibfnamefont {J.}~\bibnamefont {Wang}}, \bibinfo {author} {\bibfnamefont {C.-F.}\ \bibnamefont {Li}}, \ and\ \bibinfo {author} {\bibfnamefont {G.-C.}\ \bibnamefont {Guo}},\ }\href {https://opg.optica.org/col/abstract.cfm?URI=col-21-9-092702} {\bibfield  {journal} {\bibinfo  {journal} {Chin. Opt. Lett.}\ }\textbf {\bibinfo {volume} {21}},\ \bibinfo {pages} {092702} (\bibinfo {year} {2023})}\BibitemShut {NoStop}%
\bibitem [{\citenamefont {Duan}\ \emph {et~al.}(2001)\citenamefont {Duan}, \citenamefont {Lukin}, \citenamefont {Cirac},\ and\ \citenamefont {Zoller}}]{Duan2001}%
  \BibitemOpen
  \bibfield  {author} {\bibinfo {author} {\bibfnamefont {L.~M.}\ \bibnamefont {Duan}}, \bibinfo {author} {\bibfnamefont {M.~D.}\ \bibnamefont {Lukin}}, \bibinfo {author} {\bibfnamefont {J.~I.}\ \bibnamefont {Cirac}}, \ and\ \bibinfo {author} {\bibfnamefont {P.}~\bibnamefont {Zoller}},\ }\href {\doibase 10.1038/35106500} {\bibfield  {journal} {\bibinfo  {journal} {Nature}\ }\textbf {\bibinfo {volume} {414}},\ \bibinfo {pages} {413} (\bibinfo {year} {2001})}\BibitemShut {NoStop}%
\bibitem [{\citenamefont {Sangouard}\ \emph {et~al.}(2011)\citenamefont {Sangouard}, \citenamefont {Simon}, \citenamefont {de~Riedmatten},\ and\ \citenamefont {Gisin}}]{Sangouard2011}%
  \BibitemOpen
  \bibfield  {author} {\bibinfo {author} {\bibfnamefont {N.}~\bibnamefont {Sangouard}}, \bibinfo {author} {\bibfnamefont {C.}~\bibnamefont {Simon}}, \bibinfo {author} {\bibfnamefont {H.}~\bibnamefont {de~Riedmatten}}, \ and\ \bibinfo {author} {\bibfnamefont {N.}~\bibnamefont {Gisin}},\ }\href {\doibase 10.1103/RevModPhys.83.33} {\bibfield  {journal} {\bibinfo  {journal} {Rev. Mod. Phys.}\ }\textbf {\bibinfo {volume} {83}},\ \bibinfo {pages} {33} (\bibinfo {year} {2011})}\BibitemShut {NoStop}%
\bibitem [{\citenamefont {Higgins}\ \emph {et~al.}(2014)\citenamefont {Higgins}, \citenamefont {Benjamin}, \citenamefont {Stace}, \citenamefont {Milburn}, \citenamefont {Lovett},\ and\ \citenamefont {Gauger}}]{Higgins2014}%
  \BibitemOpen
  \bibfield  {author} {\bibinfo {author} {\bibfnamefont {K.~D.~B.}\ \bibnamefont {Higgins}}, \bibinfo {author} {\bibfnamefont {S.~C.}\ \bibnamefont {Benjamin}}, \bibinfo {author} {\bibfnamefont {T.~M.}\ \bibnamefont {Stace}}, \bibinfo {author} {\bibfnamefont {G.~J.}\ \bibnamefont {Milburn}}, \bibinfo {author} {\bibfnamefont {B.~W.}\ \bibnamefont {Lovett}}, \ and\ \bibinfo {author} {\bibfnamefont {E.~M.}\ \bibnamefont {Gauger}},\ }\href {\doibase 10.1038/ncomms5705} {\bibfield  {journal} {\bibinfo  {journal} {Nature Communications}\ }\textbf {\bibinfo {volume} {5}},\ \bibinfo {pages} {4705} (\bibinfo {year} {2014})}\BibitemShut {NoStop}%
\bibitem [{\citenamefont {Yang}\ \emph {et~al.}(2021)\citenamefont {Yang}, \citenamefont {Oh}, \citenamefont {Han}, \citenamefont {Son}, \citenamefont {Kim}, \citenamefont {Kim}, \citenamefont {Lee},\ and\ \citenamefont {An}}]{Yang2021}%
  \BibitemOpen
  \bibfield  {author} {\bibinfo {author} {\bibfnamefont {D.}~\bibnamefont {Yang}}, \bibinfo {author} {\bibfnamefont {S.-h.}\ \bibnamefont {Oh}}, \bibinfo {author} {\bibfnamefont {J.}~\bibnamefont {Han}}, \bibinfo {author} {\bibfnamefont {G.}~\bibnamefont {Son}}, \bibinfo {author} {\bibfnamefont {J.}~\bibnamefont {Kim}}, \bibinfo {author} {\bibfnamefont {J.}~\bibnamefont {Kim}}, \bibinfo {author} {\bibfnamefont {M.}~\bibnamefont {Lee}}, \ and\ \bibinfo {author} {\bibfnamefont {K.}~\bibnamefont {An}},\ }\href {\doibase 10.1038/s41566-021-00770-6} {\bibfield  {journal} {\bibinfo  {journal} {Nature Photonics}\ }\textbf {\bibinfo {volume} {15}},\ \bibinfo {pages} {272} (\bibinfo {year} {2021})}\BibitemShut {NoStop}%
\bibitem [{\citenamefont {Shen}\ and\ \citenamefont {Fan}(2005)}]{Shen05}%
  \BibitemOpen
  \bibfield  {author} {\bibinfo {author} {\bibfnamefont {J.~T.}\ \bibnamefont {Shen}}\ and\ \bibinfo {author} {\bibfnamefont {S.}~\bibnamefont {Fan}},\ }\href {\doibase 10.1364/OL.30.002001} {\bibfield  {journal} {\bibinfo  {journal} {Opt. Lett.}\ }\textbf {\bibinfo {volume} {30}},\ \bibinfo {pages} {2001} (\bibinfo {year} {2005})}\BibitemShut {NoStop}%
\bibitem [{\citenamefont {Weidem\"uller}\ \emph {et~al.}(1995)\citenamefont {Weidem\"uller}, \citenamefont {Hemmerich}, \citenamefont {G\"orlitz}, \citenamefont {Esslinger},\ and\ \citenamefont {H\"ansch}}]{Weidemuller1995}%
  \BibitemOpen
  \bibfield  {author} {\bibinfo {author} {\bibfnamefont {M.}~\bibnamefont {Weidem\"uller}}, \bibinfo {author} {\bibfnamefont {A.}~\bibnamefont {Hemmerich}}, \bibinfo {author} {\bibfnamefont {A.}~\bibnamefont {G\"orlitz}}, \bibinfo {author} {\bibfnamefont {T.}~\bibnamefont {Esslinger}}, \ and\ \bibinfo {author} {\bibfnamefont {T.~W.}\ \bibnamefont {H\"ansch}},\ }\href {\doibase 10.1103/PhysRevLett.75.4583} {\bibfield  {journal} {\bibinfo  {journal} {Phys. Rev. Lett.}\ }\textbf {\bibinfo {volume} {75}},\ \bibinfo {pages} {4583} (\bibinfo {year} {1995})}\BibitemShut {NoStop}%
\bibitem [{\citenamefont {Birkl}\ \emph {et~al.}(1995)\citenamefont {Birkl}, \citenamefont {Gatzke}, \citenamefont {Deutsch}, \citenamefont {Rolston},\ and\ \citenamefont {Phillips}}]{Birkl1995}%
  \BibitemOpen
  \bibfield  {author} {\bibinfo {author} {\bibfnamefont {G.}~\bibnamefont {Birkl}}, \bibinfo {author} {\bibfnamefont {M.}~\bibnamefont {Gatzke}}, \bibinfo {author} {\bibfnamefont {I.~H.}\ \bibnamefont {Deutsch}}, \bibinfo {author} {\bibfnamefont {S.~L.}\ \bibnamefont {Rolston}}, \ and\ \bibinfo {author} {\bibfnamefont {W.~D.}\ \bibnamefont {Phillips}},\ }\href {\doibase 10.1103/PhysRevLett.75.2823} {\bibfield  {journal} {\bibinfo  {journal} {Phys. Rev. Lett.}\ }\textbf {\bibinfo {volume} {75}},\ \bibinfo {pages} {2823} (\bibinfo {year} {1995})}\BibitemShut {NoStop}%
\bibitem [{\citenamefont {Corzo}\ \emph {et~al.}(2016)\citenamefont {Corzo}, \citenamefont {Gouraud}, \citenamefont {Chandra}, \citenamefont {Goban}, \citenamefont {Sheremet}, \citenamefont {Kupriyanov},\ and\ \citenamefont {Laurat}}]{Corzo2016}%
  \BibitemOpen
  \bibfield  {author} {\bibinfo {author} {\bibfnamefont {N.~V.}\ \bibnamefont {Corzo}}, \bibinfo {author} {\bibfnamefont {B.}~\bibnamefont {Gouraud}}, \bibinfo {author} {\bibfnamefont {A.}~\bibnamefont {Chandra}}, \bibinfo {author} {\bibfnamefont {A.}~\bibnamefont {Goban}}, \bibinfo {author} {\bibfnamefont {A.~S.}\ \bibnamefont {Sheremet}}, \bibinfo {author} {\bibfnamefont {D.~V.}\ \bibnamefont {Kupriyanov}}, \ and\ \bibinfo {author} {\bibfnamefont {J.}~\bibnamefont {Laurat}},\ }\href {\doibase 10.1103/PhysRevLett.117.133603} {\bibfield  {journal} {\bibinfo  {journal} {Phys. Rev. Lett.}\ }\textbf {\bibinfo {volume} {117}},\ \bibinfo {pages} {133603} (\bibinfo {year} {2016})}\BibitemShut {NoStop}%
\bibitem [{\citenamefont {S\o{}rensen}\ \emph {et~al.}(2016)\citenamefont {S\o{}rensen}, \citenamefont {B\'eguin}, \citenamefont {Kluge}, \citenamefont {Iakoupov}, \citenamefont {S\o{}rensen}, \citenamefont {M\"uller}, \citenamefont {Polzik},\ and\ \citenamefont {Appel}}]{Sorensen2016}%
  \BibitemOpen
  \bibfield  {author} {\bibinfo {author} {\bibfnamefont {H.~L.}\ \bibnamefont {S\o{}rensen}}, \bibinfo {author} {\bibfnamefont {J.-B.}\ \bibnamefont {B\'eguin}}, \bibinfo {author} {\bibfnamefont {K.~W.}\ \bibnamefont {Kluge}}, \bibinfo {author} {\bibfnamefont {I.}~\bibnamefont {Iakoupov}}, \bibinfo {author} {\bibfnamefont {A.~S.}\ \bibnamefont {S\o{}rensen}}, \bibinfo {author} {\bibfnamefont {J.~H.}\ \bibnamefont {M\"uller}}, \bibinfo {author} {\bibfnamefont {E.~S.}\ \bibnamefont {Polzik}}, \ and\ \bibinfo {author} {\bibfnamefont {J.}~\bibnamefont {Appel}},\ }\href {\doibase 10.1103/PhysRevLett.117.133604} {\bibfield  {journal} {\bibinfo  {journal} {Phys. Rev. Lett.}\ }\textbf {\bibinfo {volume} {117}},\ \bibinfo {pages} {133604} (\bibinfo {year} {2016})}\BibitemShut {NoStop}%
\bibitem [{\citenamefont {Guimond}\ \emph {et~al.}(2016)\citenamefont {Guimond}, \citenamefont {Roulet}, \citenamefont {Le},\ and\ \citenamefont {Scarani}}]{Guimond2016}%
  \BibitemOpen
  \bibfield  {author} {\bibinfo {author} {\bibfnamefont {P.-O.}\ \bibnamefont {Guimond}}, \bibinfo {author} {\bibfnamefont {A.}~\bibnamefont {Roulet}}, \bibinfo {author} {\bibfnamefont {H.~N.}\ \bibnamefont {Le}}, \ and\ \bibinfo {author} {\bibfnamefont {V.}~\bibnamefont {Scarani}},\ }\href {\doibase 10.1103/PhysRevA.93.023808} {\bibfield  {journal} {\bibinfo  {journal} {Phys. Rev. A}\ }\textbf {\bibinfo {volume} {93}},\ \bibinfo {pages} {023808} (\bibinfo {year} {2016})}\BibitemShut {NoStop}%
\bibitem [{\citenamefont {Sheremet}\ \emph {et~al.}(2023)\citenamefont {Sheremet}, \citenamefont {Petrov}, \citenamefont {Iorsh}, \citenamefont {Poshakinskiy},\ and\ \citenamefont {Poddubny}}]{Sheremet23}%
  \BibitemOpen
  \bibfield  {author} {\bibinfo {author} {\bibfnamefont {A.~S.}\ \bibnamefont {Sheremet}}, \bibinfo {author} {\bibfnamefont {M.~I.}\ \bibnamefont {Petrov}}, \bibinfo {author} {\bibfnamefont {I.~V.}\ \bibnamefont {Iorsh}}, \bibinfo {author} {\bibfnamefont {A.~V.}\ \bibnamefont {Poshakinskiy}}, \ and\ \bibinfo {author} {\bibfnamefont {A.~N.}\ \bibnamefont {Poddubny}},\ }\href {\doibase 10.1103/RevModPhys.95.015002} {\bibfield  {journal} {\bibinfo  {journal} {Rev. Mod. Phys.}\ }\textbf {\bibinfo {volume} {95}},\ \bibinfo {pages} {015002} (\bibinfo {year} {2023})}\BibitemShut {NoStop}%
\bibitem [{\citenamefont {Giessen}\ \emph {et~al.}(1996)\citenamefont {Giessen}, \citenamefont {Berger}, \citenamefont {Mohs}, \citenamefont {Meystre},\ and\ \citenamefont {Yelin}}]{Giessen1996}%
  \BibitemOpen
  \bibfield  {author} {\bibinfo {author} {\bibfnamefont {H.}~\bibnamefont {Giessen}}, \bibinfo {author} {\bibfnamefont {J.~D.}\ \bibnamefont {Berger}}, \bibinfo {author} {\bibfnamefont {G.}~\bibnamefont {Mohs}}, \bibinfo {author} {\bibfnamefont {P.}~\bibnamefont {Meystre}}, \ and\ \bibinfo {author} {\bibfnamefont {S.~F.}\ \bibnamefont {Yelin}},\ }\href {\doibase 10.1103/PhysRevA.53.2816} {\bibfield  {journal} {\bibinfo  {journal} {Phys. Rev. A}\ }\textbf {\bibinfo {volume} {53}},\ \bibinfo {pages} {2816} (\bibinfo {year} {1996})}\BibitemShut {NoStop}%
\bibitem [{\citenamefont {Krimer}\ \emph {et~al.}(2014)\citenamefont {Krimer}, \citenamefont {Liertzer}, \citenamefont {Rotter},\ and\ \citenamefont {T\"ureci}}]{Krimer2014}%
  \BibitemOpen
  \bibfield  {author} {\bibinfo {author} {\bibfnamefont {D.~O.}\ \bibnamefont {Krimer}}, \bibinfo {author} {\bibfnamefont {M.}~\bibnamefont {Liertzer}}, \bibinfo {author} {\bibfnamefont {S.}~\bibnamefont {Rotter}}, \ and\ \bibinfo {author} {\bibfnamefont {H.~E.}\ \bibnamefont {T\"ureci}},\ }\href {\doibase 10.1103/PhysRevA.89.033820} {\bibfield  {journal} {\bibinfo  {journal} {Phys. Rev. A}\ }\textbf {\bibinfo {volume} {89}},\ \bibinfo {pages} {033820} (\bibinfo {year} {2014})}\BibitemShut {NoStop}%
\bibitem [{\citenamefont {Sinha}\ \emph {et~al.}(2020{\natexlab{a}})\citenamefont {Sinha}, \citenamefont {Meystre}, \citenamefont {Goldschmidt}, \citenamefont {Fatemi}, \citenamefont {Rolston},\ and\ \citenamefont {Solano}}]{Sinha20a}%
  \BibitemOpen
  \bibfield  {author} {\bibinfo {author} {\bibfnamefont {K.}~\bibnamefont {Sinha}}, \bibinfo {author} {\bibfnamefont {P.}~\bibnamefont {Meystre}}, \bibinfo {author} {\bibfnamefont {E.~A.}\ \bibnamefont {Goldschmidt}}, \bibinfo {author} {\bibfnamefont {F.~K.}\ \bibnamefont {Fatemi}}, \bibinfo {author} {\bibfnamefont {S.~L.}\ \bibnamefont {Rolston}}, \ and\ \bibinfo {author} {\bibfnamefont {P.}~\bibnamefont {Solano}},\ }\href {\doibase 10.1103/PhysRevLett.124.043603} {\bibfield  {journal} {\bibinfo  {journal} {Phys. Rev. Lett.}\ }\textbf {\bibinfo {volume} {124}},\ \bibinfo {pages} {043603} (\bibinfo {year} {2020}{\natexlab{a}})}\BibitemShut {NoStop}%
\bibitem [{\citenamefont {Sinha}\ \emph {et~al.}(2020{\natexlab{b}})\citenamefont {Sinha}, \citenamefont {Gonz\'alez-Tudela}, \citenamefont {Lu},\ and\ \citenamefont {Solano}}]{Sinha20b}%
  \BibitemOpen
  \bibfield  {author} {\bibinfo {author} {\bibfnamefont {K.}~\bibnamefont {Sinha}}, \bibinfo {author} {\bibfnamefont {A.}~\bibnamefont {Gonz\'alez-Tudela}}, \bibinfo {author} {\bibfnamefont {Y.}~\bibnamefont {Lu}}, \ and\ \bibinfo {author} {\bibfnamefont {P.}~\bibnamefont {Solano}},\ }\href {\doibase 10.1103/PhysRevA.102.043718} {\bibfield  {journal} {\bibinfo  {journal} {Phys. Rev. A}\ }\textbf {\bibinfo {volume} {102}},\ \bibinfo {pages} {043718} (\bibinfo {year} {2020}{\natexlab{b}})}\BibitemShut {NoStop}%
\bibitem [{\citenamefont {Crowder}\ \emph {et~al.}(2024)\citenamefont {Crowder}, \citenamefont {Ramunno},\ and\ \citenamefont {Hughes}}]{Crowder2024}%
  \BibitemOpen
  \bibfield  {author} {\bibinfo {author} {\bibfnamefont {G.}~\bibnamefont {Crowder}}, \bibinfo {author} {\bibfnamefont {L.}~\bibnamefont {Ramunno}}, \ and\ \bibinfo {author} {\bibfnamefont {S.}~\bibnamefont {Hughes}},\ }\href {\doibase 10.1103/PhysRevA.110.L031703} {\bibfield  {journal} {\bibinfo  {journal} {Phys. Rev. A}\ }\textbf {\bibinfo {volume} {110}},\ \bibinfo {pages} {L031703} (\bibinfo {year} {2024})}\BibitemShut {NoStop}%
\bibitem [{\citenamefont {Zheng}\ and\ \citenamefont {Baranger}(2013)}]{Zheng}%
  \BibitemOpen
  \bibfield  {author} {\bibinfo {author} {\bibfnamefont {H.}~\bibnamefont {Zheng}}\ and\ \bibinfo {author} {\bibfnamefont {H.~U.}\ \bibnamefont {Baranger}},\ }\href {\doibase 10.1103/PhysRevLett.110.113601} {\bibfield  {journal} {\bibinfo  {journal} {Phys. Rev. Lett.}\ }\textbf {\bibinfo {volume} {110}},\ \bibinfo {pages} {113601} (\bibinfo {year} {2013})}\BibitemShut {NoStop}%
\bibitem [{\citenamefont {Alvarez-Giron}\ \emph {et~al.}(2024)\citenamefont {Alvarez-Giron}, \citenamefont {Solano}, \citenamefont {Sinha},\ and\ \citenamefont {Barberis-Blostein}}]{Giron}%
  \BibitemOpen
  \bibfield  {author} {\bibinfo {author} {\bibfnamefont {W.}~\bibnamefont {Alvarez-Giron}}, \bibinfo {author} {\bibfnamefont {P.}~\bibnamefont {Solano}}, \bibinfo {author} {\bibfnamefont {K.}~\bibnamefont {Sinha}}, \ and\ \bibinfo {author} {\bibfnamefont {P.}~\bibnamefont {Barberis-Blostein}},\ }\href {\doibase 10.1103/PhysRevResearch.6.023213} {\bibfield  {journal} {\bibinfo  {journal} {Phys. Rev. Res.}\ }\textbf {\bibinfo {volume} {6}},\ \bibinfo {pages} {023213} (\bibinfo {year} {2024})}\BibitemShut {NoStop}%
\bibitem [{\citenamefont {Dorner}\ and\ \citenamefont {Zoller}(2002)}]{DornerZoller}%
  \BibitemOpen
  \bibfield  {author} {\bibinfo {author} {\bibfnamefont {U.}~\bibnamefont {Dorner}}\ and\ \bibinfo {author} {\bibfnamefont {P.}~\bibnamefont {Zoller}},\ }\href {\doibase 10.1103/PhysRevA.66.023816} {\bibfield  {journal} {\bibinfo  {journal} {Phys. Rev. A}\ }\textbf {\bibinfo {volume} {66}},\ \bibinfo {pages} {023816} (\bibinfo {year} {2002})}\BibitemShut {NoStop}%
\bibitem [{\citenamefont {Dinc}\ and\ \citenamefont {Bra\ifmmode~\acute{n}\else \'{n}\fi{}czyk}(2019)}]{Dinc2019}%
  \BibitemOpen
  \bibfield  {author} {\bibinfo {author} {\bibfnamefont {F.}~\bibnamefont {Dinc}}\ and\ \bibinfo {author} {\bibfnamefont {A.~M.}\ \bibnamefont {Bra\ifmmode~\acute{n}\else \'{n}\fi{}czyk}},\ }\href {\doibase 10.1103/PhysRevResearch.1.032042} {\bibfield  {journal} {\bibinfo  {journal} {Phys. Rev. Res.}\ }\textbf {\bibinfo {volume} {1}},\ \bibinfo {pages} {032042} (\bibinfo {year} {2019})}\BibitemShut {NoStop}%
\bibitem [{\citenamefont {Blow}\ \emph {et~al.}(1990)\citenamefont {Blow}, \citenamefont {Loudon}, \citenamefont {Phoenix},\ and\ \citenamefont {Shepherd}}]{Blow}%
  \BibitemOpen
  \bibfield  {author} {\bibinfo {author} {\bibfnamefont {K.~J.}\ \bibnamefont {Blow}}, \bibinfo {author} {\bibfnamefont {R.}~\bibnamefont {Loudon}}, \bibinfo {author} {\bibfnamefont {S.~J.~D.}\ \bibnamefont {Phoenix}}, \ and\ \bibinfo {author} {\bibfnamefont {T.~J.}\ \bibnamefont {Shepherd}},\ }\href {\doibase 10.1103/PhysRevA.42.4102} {\bibfield  {journal} {\bibinfo  {journal} {Phys. Rev. A}\ }\textbf {\bibinfo {volume} {42}},\ \bibinfo {pages} {4102} (\bibinfo {year} {1990})}\BibitemShut {NoStop}%
\bibitem [{\citenamefont {Meystre}\ and\ \citenamefont {Sargent}(2007)}]{MeystreBook2007}%
  \BibitemOpen
  \bibfield  {author} {\bibinfo {author} {\bibfnamefont {P.}~\bibnamefont {Meystre}}\ and\ \bibinfo {author} {\bibfnamefont {M.}~\bibnamefont {Sargent}},\ }\href@noop {} {\emph {\bibinfo {title} {Elements of quantum optics}}}\ (\bibinfo  {publisher} {Springer Berlin Heidelberg},\ \bibinfo {year} {2007})\BibitemShut {NoStop}%
\bibitem [{\citenamefont {Sinha}\ \emph {et~al.}(2024)\citenamefont {Sinha}, \citenamefont {Parra-Contreras}, \citenamefont {Das},\ and\ \citenamefont {Solano}}]{Sinha2024}%
  \BibitemOpen
  \bibfield  {author} {\bibinfo {author} {\bibfnamefont {K.}~\bibnamefont {Sinha}}, \bibinfo {author} {\bibfnamefont {J.}~\bibnamefont {Parra-Contreras}}, \bibinfo {author} {\bibfnamefont {A.}~\bibnamefont {Das}}, \ and\ \bibinfo {author} {\bibfnamefont {P.}~\bibnamefont {Solano}},\ }\href {https://arxiv.org/abs/2402.10303} {\bibfield  {journal} {\bibinfo  {journal} {arXiv}\ } (\bibinfo {year} {2024})},\ \Eprint {http://arxiv.org/abs/2402.10303} {2402.10303 [quant-ph]} \BibitemShut {NoStop}%
\bibitem [{\citenamefont {Agarwal}(1984)}]{Agarwal1984}%
  \BibitemOpen
  \bibfield  {author} {\bibinfo {author} {\bibfnamefont {G.~S.}\ \bibnamefont {Agarwal}},\ }\href {\doibase 10.1103/PhysRevLett.53.1732} {\bibfield  {journal} {\bibinfo  {journal} {Phys. Rev. Lett.}\ }\textbf {\bibinfo {volume} {53}},\ \bibinfo {pages} {1732} (\bibinfo {year} {1984})}\BibitemShut {NoStop}%
\bibitem [{\citenamefont {Meiser}\ and\ \citenamefont {Meystre}(2006)}]{Meiser2006}%
  \BibitemOpen
  \bibfield  {author} {\bibinfo {author} {\bibfnamefont {D.}~\bibnamefont {Meiser}}\ and\ \bibinfo {author} {\bibfnamefont {P.}~\bibnamefont {Meystre}},\ }\href {\doibase 10.1103/PhysRevA.74.065801} {\bibfield  {journal} {\bibinfo  {journal} {Phys. Rev. A}\ }\textbf {\bibinfo {volume} {74}},\ \bibinfo {pages} {065801} (\bibinfo {year} {2006})}\BibitemShut {NoStop}%
\bibitem [{\citenamefont {Johnson}\ \emph {et~al.}(2019)\citenamefont {Johnson}, \citenamefont {Blaha}, \citenamefont {Ulanov}, \citenamefont {Rauschenbeutel}, \citenamefont {Schneeweiss},\ and\ \citenamefont {Volz}}]{Johnson2019}%
  \BibitemOpen
  \bibfield  {author} {\bibinfo {author} {\bibfnamefont {A.}~\bibnamefont {Johnson}}, \bibinfo {author} {\bibfnamefont {M.}~\bibnamefont {Blaha}}, \bibinfo {author} {\bibfnamefont {A.~E.}\ \bibnamefont {Ulanov}}, \bibinfo {author} {\bibfnamefont {A.}~\bibnamefont {Rauschenbeutel}}, \bibinfo {author} {\bibfnamefont {P.}~\bibnamefont {Schneeweiss}}, \ and\ \bibinfo {author} {\bibfnamefont {J.}~\bibnamefont {Volz}},\ }\href {\doibase 10.1103/PhysRevLett.123.243602} {\bibfield  {journal} {\bibinfo  {journal} {Phys. Rev. Lett.}\ }\textbf {\bibinfo {volume} {123}},\ \bibinfo {pages} {243602} (\bibinfo {year} {2019})}\BibitemShut {NoStop}%
\bibitem [{\citenamefont {Blaha}\ \emph {et~al.}(2022)\citenamefont {Blaha}, \citenamefont {Johnson}, \citenamefont {Rauschenbeutel},\ and\ \citenamefont {Volz}}]{Blaha2022}%
  \BibitemOpen
  \bibfield  {author} {\bibinfo {author} {\bibfnamefont {M.}~\bibnamefont {Blaha}}, \bibinfo {author} {\bibfnamefont {A.}~\bibnamefont {Johnson}}, \bibinfo {author} {\bibfnamefont {A.}~\bibnamefont {Rauschenbeutel}}, \ and\ \bibinfo {author} {\bibfnamefont {J.}~\bibnamefont {Volz}},\ }\href {\doibase 10.1103/PhysRevA.105.013719} {\bibfield  {journal} {\bibinfo  {journal} {Phys. Rev. A}\ }\textbf {\bibinfo {volume} {105}},\ \bibinfo {pages} {013719} (\bibinfo {year} {2022})}\BibitemShut {NoStop}%
\bibitem [{\citenamefont {Calaj\'o}\ \emph {et~al.}(2019)\citenamefont {Calaj\'o}, \citenamefont {Fang}, \citenamefont {Baranger},\ and\ \citenamefont {Ciccarello}}]{Calajo}%
  \BibitemOpen
  \bibfield  {author} {\bibinfo {author} {\bibfnamefont {G.}~\bibnamefont {Calaj\'o}}, \bibinfo {author} {\bibfnamefont {Y.-L.~L.}\ \bibnamefont {Fang}}, \bibinfo {author} {\bibfnamefont {H.~U.}\ \bibnamefont {Baranger}}, \ and\ \bibinfo {author} {\bibfnamefont {F.}~\bibnamefont {Ciccarello}},\ }\href {\doibase 10.1103/PhysRevLett.122.073601} {\bibfield  {journal} {\bibinfo  {journal} {Phys. Rev. Lett.}\ }\textbf {\bibinfo {volume} {122}},\ \bibinfo {pages} {073601} (\bibinfo {year} {2019})}\BibitemShut {NoStop}%
\bibitem [{\citenamefont {Guo}\ \emph {et~al.}(2020)\citenamefont {Guo}, \citenamefont {Kockum}, \citenamefont {Marquardt},\ and\ \citenamefont {Johansson}}]{Guo2020}%
  \BibitemOpen
  \bibfield  {author} {\bibinfo {author} {\bibfnamefont {L.}~\bibnamefont {Guo}}, \bibinfo {author} {\bibfnamefont {A.~F.}\ \bibnamefont {Kockum}}, \bibinfo {author} {\bibfnamefont {F.}~\bibnamefont {Marquardt}}, \ and\ \bibinfo {author} {\bibfnamefont {G.}~\bibnamefont {Johansson}},\ }\href {\doibase 10.1103/PhysRevResearch.2.043014} {\bibfield  {journal} {\bibinfo  {journal} {Phys. Rev. Res.}\ }\textbf {\bibinfo {volume} {2}},\ \bibinfo {pages} {043014} (\bibinfo {year} {2020})}\BibitemShut {NoStop}%
\bibitem [{\citenamefont {Trivedi}\ \emph {et~al.}(2021)\citenamefont {Trivedi}, \citenamefont {Malz}, \citenamefont {Sun}, \citenamefont {Fan},\ and\ \citenamefont {Vu\ifmmode \check{c}\else \v{c}\fi{}kovi\ifmmode~\acute{c}\else \'{c}\fi{}}}]{Trivedi2021}%
  \BibitemOpen
  \bibfield  {author} {\bibinfo {author} {\bibfnamefont {R.}~\bibnamefont {Trivedi}}, \bibinfo {author} {\bibfnamefont {D.}~\bibnamefont {Malz}}, \bibinfo {author} {\bibfnamefont {S.}~\bibnamefont {Sun}}, \bibinfo {author} {\bibfnamefont {S.}~\bibnamefont {Fan}}, \ and\ \bibinfo {author} {\bibfnamefont {J.}~\bibnamefont {Vu\ifmmode \check{c}\else \v{c}\fi{}kovi\ifmmode~\acute{c}\else \'{c}\fi{}}},\ }\href {\doibase 10.1103/PhysRevA.104.013705} {\bibfield  {journal} {\bibinfo  {journal} {Phys. Rev. A}\ }\textbf {\bibinfo {volume} {104}},\ \bibinfo {pages} {013705} (\bibinfo {year} {2021})}\BibitemShut {NoStop}%
\bibitem [{\citenamefont {Mirhosseini}\ \emph {et~al.}(2019)\citenamefont {Mirhosseini}, \citenamefont {Kim}, \citenamefont {Zhang}, \citenamefont {Sipahigil}, \citenamefont {Dieterle}, \citenamefont {Keller}, \citenamefont {Asenjo-Garcia}, \citenamefont {Chang},\ and\ \citenamefont {Painter}}]{Mirhosseini2019}%
  \BibitemOpen
  \bibfield  {author} {\bibinfo {author} {\bibfnamefont {M.}~\bibnamefont {Mirhosseini}}, \bibinfo {author} {\bibfnamefont {E.}~\bibnamefont {Kim}}, \bibinfo {author} {\bibfnamefont {X.}~\bibnamefont {Zhang}}, \bibinfo {author} {\bibfnamefont {A.}~\bibnamefont {Sipahigil}}, \bibinfo {author} {\bibfnamefont {P.~B.}\ \bibnamefont {Dieterle}}, \bibinfo {author} {\bibfnamefont {A.~J.}\ \bibnamefont {Keller}}, \bibinfo {author} {\bibfnamefont {A.}~\bibnamefont {Asenjo-Garcia}}, \bibinfo {author} {\bibfnamefont {D.~E.}\ \bibnamefont {Chang}}, \ and\ \bibinfo {author} {\bibfnamefont {O.}~\bibnamefont {Painter}},\ }\href {\doibase 10.1038/s41586-019-1196-1} {\bibfield  {journal} {\bibinfo  {journal} {Nature}\ }\textbf {\bibinfo {volume} {569}},\ \bibinfo {pages} {692} (\bibinfo {year} {2019})}\BibitemShut {NoStop}%
\bibitem [{\citenamefont {Kim}\ \emph {et~al.}(2025)\citenamefont {Kim}, \citenamefont {Lanuza},\ and\ \citenamefont {Schneble}}]{Kim2025}%
  \BibitemOpen
  \bibfield  {author} {\bibinfo {author} {\bibfnamefont {Y.}~\bibnamefont {Kim}}, \bibinfo {author} {\bibfnamefont {A.}~\bibnamefont {Lanuza}}, \ and\ \bibinfo {author} {\bibfnamefont {D.}~\bibnamefont {Schneble}},\ }\href {\doibase 10.1038/s41567-024-02676-w} {\bibfield  {journal} {\bibinfo  {journal} {Nature Physics}\ }\textbf {\bibinfo {volume} {21}},\ \bibinfo {pages} {70} (\bibinfo {year} {2025})}\BibitemShut {NoStop}%
\bibitem [{\citenamefont {Keefe}\ \emph {et~al.}(2024)\citenamefont {Keefe}, \citenamefont {Agarwal},\ and\ \citenamefont {Kamal}}]{keefe2024}%
  \BibitemOpen
  \bibfield  {author} {\bibinfo {author} {\bibfnamefont {A.}~\bibnamefont {Keefe}}, \bibinfo {author} {\bibfnamefont {N.}~\bibnamefont {Agarwal}}, \ and\ \bibinfo {author} {\bibfnamefont {A.}~\bibnamefont {Kamal}},\ }\href {https://arxiv.org/abs/2405.01722} {\bibfield  {journal} {\bibinfo  {journal} {arXiv:2405.01722}\ } (\bibinfo {year} {2024})}\BibitemShut {NoStop}%
\bibitem [{\citenamefont {Cilluffo}\ \emph {et~al.}(2024)\citenamefont {Cilluffo}, \citenamefont {Ferialdi}, \citenamefont {Palma}, \citenamefont {Calaj\`{o}},\ and\ \citenamefont {Ciccarello}}]{cilluffo2024}%
  \BibitemOpen
  \bibfield  {author} {\bibinfo {author} {\bibfnamefont {D.}~\bibnamefont {Cilluffo}}, \bibinfo {author} {\bibfnamefont {L.}~\bibnamefont {Ferialdi}}, \bibinfo {author} {\bibfnamefont {G.~M.}\ \bibnamefont {Palma}}, \bibinfo {author} {\bibfnamefont {G.}~\bibnamefont {Calaj\`{o}}}, \ and\ \bibinfo {author} {\bibfnamefont {F.}~\bibnamefont {Ciccarello}},\ }\href {https://arxiv.org/abs/2403.07110} {\bibfield  {journal} {\bibinfo  {journal} {arXiv:2403.07110}\ } (\bibinfo {year} {2024})}\BibitemShut {NoStop}%
\bibitem [{\citenamefont {Abramowitz}\ and\ \citenamefont {Stegun}(1964)}]{stegun}%
  \BibitemOpen
  \bibfield  {author} {\bibinfo {author} {\bibfnamefont {M.}~\bibnamefont {Abramowitz}}\ and\ \bibinfo {author} {\bibfnamefont {I.~A.}\ \bibnamefont {Stegun}},\ }\href@noop {} {\emph {\bibinfo {title} {Handbook of Mathematical Functions with Formulas, Graphs, and Mathematical Tables}}},\ \bibinfo {edition} {ninth dover printing, tenth gpo printing}\ ed.\ (\bibinfo  {publisher} {Dover},\ \bibinfo {address} {New York},\ \bibinfo {year} {1964})\BibitemShut {NoStop}%
\end{thebibliography}%
\end{document}